\documentclass[
superscriptaddress,
pra,
twocolumn,
]{revtex4-2}

\usepackage{graphicx}
\usepackage{dcolumn}
\usepackage{bm}
\usepackage{soul}

\usepackage{mathtools}
\usepackage{braket}
\usepackage{ mathrsfs}
\usepackage{physics}
\usepackage{xcolor}
\usepackage{dsfont}
\usepackage{afterpage}
\usepackage{soul}

\begin{document}

\preprint{APS/123-QED}
\title{Extreme-ultraviolet spectroscopy using quantum logic: \\[0.5em] a feasibility study for the $1S-2S$ transition in singly-ionized helium}

\author{A. Mart\'inez de Velasco}
\author{V.\,P.\,J. Barb\'e}
\author{E.\,L. Gr\"undeman}
\author{A. D\'iaz Calder\'on}
\author{M. Collombon}
\author{J.\,J. Krauth}
\author{C.\,F. Roth}
\author{M. Favier}
\affiliation{Vrije Universiteit Amsterdam, De Boelelaan 1100, 1081 HZ Amsterdam, Netherlands}
\author{R. Taieb}
\affiliation{Sorbonne Universit\'e, CNRS, Laboratoire de Chimie Physique-Mati´ere et Rayonnement, LCPMR, F-75005 Paris Cedex 05, France}
\author{T. E. Mehlst\"aubler}
\author{P.O. Schmidt}
\affiliation{Physikalisch-Technische Bundesanstalt, Byndesallee 100, 38116 Braunschweig, Germany} \affiliation{Institut f\"ur Quantenoptik, Leibniz Universit\"at Hannover, Welfengarten 1, 30167, Hannover, Germany}
\author{L.\,S. Dreissen}
\author{K.\,S.\,E. Eikema}
\affiliation{Vrije Universiteit Amsterdam, De Boelelaan 1100, 1081 HZ Amsterdam, Netherlands}

\date{\today}

\begin{abstract}
Extreme-ultraviolet (XUV) spectroscopy represents an important new direction in precision physics, with potential applications ranging from the metrology of fundamental constants to tests of physics beyond the Standard Model. However, the application of quantum control methods for precision spectroscopy remains an open challenge in the XUV range. Here we present a novel quantum logic (QL) spectroscopy method for precision spectroscopy of weak XUV transitions, and numerically validate its feasibility for the $1S-2S$ transition at 40.81\,eV in singly-ionized helium (He$^{+}$). We propose a scheme based on a single He$^{+}$ ion co-trapped with a Be$^{+}$ ion in a Paul trap, and He$^{+}$ excitation with pairs of frequency-comb (FC) laser pulses upconverted to the XUV via High-Harmonic Generation (HHG). We investigate a nondestructive QL scheme to detect $1S-2S$ excitation, and compare its performance with a destructive readout based on state-selective ionization. We model the phase coherence of the XUV light and show that it is compatible with precision spectroscopy, provided an optical cavity is used to filter the FC pulses prior to HHG. We model the motional excitation dynamics of trapped ions outside the Lamb-Dicke regime, and numerically validate a scheme we proposed in \cite{Grundeman} to cancel the first-order Doppler broadening and the recoil shift by synchronizing the ion's secular period with the time delay between the two excitation pulses. We also study the impact of quantum projection noise and false positive counts on the projected spectroscopic uncertainty. We show that precision spectroscopy of the $1S-2S$ transition in He$^{+}$ at the 10 kHz level is feasible within reasonable data acquisition times, which could allow for improved tests of quantum electrodynamics (QED), a measurement of the Rydberg constant $R_{\infty}$ independent of hydrogen measurements, or an improved determination of the alpha particle and helion charge radii. The proposed method may also be applied to XUV spectroscopy of other ions outside the Lamb-Dicke regime.
\end{abstract}

\maketitle

\section{Introduction}

Precision spectroscopy in the extreme ultraviolet range ($\lambda$\,=\,10\,--\,121\,nm) is of great interest for the determination of fundamental constants, tests of fundamental physics and the development of novel quantum sensing devices such as atomic and nuclear clocks~\cite{Pupeza2021,Eikema2011,HCI_review,Lyu2025}. XUV spectroscopy has advanced considerably in recent years due to the combination of optical frequency combs (OFCs) and high-harmonic generation (HHG) to produce phase-coherent light at short wavelengths, with experiments achieving fractional spectroscopic accuracies at the $10^{-10}$ level in atomic beams of noble-gas elements~\cite{Laura_xenon,Cingoz}. Bringing together these XUV techniques with single-atom,  single-quantum-state control techniques, which have enabled extremely precise spectroscopy at more conventional wavelengths~\cite{brewer_Al_2019,King2022,hb3c-dk28,King2022,ma_quantum-logic-based_2024,dawel2026alclock16times1018systematic}, is therefore very promising. However, it remains a significant challenge owing to the unique properties of XUV light and the specific requirements they impose on established quantum control methods.

In this work, we propose and numerically validate a novel quantum logic (QL) method for the spectroscopy of weak XUV transitions, based on the combination of Ramsey-Comb Spectroscopy (RCS)~\cite{Jonas, Jonas_theory} and a QL protocol for nondestructive spectroscopy readout.
We consider the case of the $1S-2S$ transition of He$^+$, a two-photon XUV transition that is of particular interest for an improved determination of the alpha particle and helion charge radii, novel precision tests of quantum electrodynamics (QED) or an alternative measurement of the Rydberg constant~\cite{krauth2019paving}.
In general, there is a drive to extend precision physics to shorter wavelengths, and a prime example of that is the recent demonstration of nuclear clocks based on a transition in the nucleus of thorium-229 at a vacuum-ultraviolet (VUV) wavelength of $\lambda$~=~148\,nm~\cite{huang2026nuclearclockbased229th,decol2026thorium229opticalnuclearclock,thorium1, thorium2,Morawetz2026,Ooi2026}. The approach we focus on for the $1S-2S$ transition in He$^{+}$ paves the way to perform precision spectroscopy deep into the XUV, with interesting other targets such as transitions in highly-charged ions, which have been proposed for novel tests of physics beyond the Standard Model~\cite{HCI_review,Lyu2025}, or the weak 76\,eV nuclear transition in uranium-235 ions ($\lambda=16$\,nm), which has been proposed as a future nuclear clock~\cite{PhysRevLett.121.253002}. 


We recently demonstrated the first excitation of the $1S-2S$ transition of He$^+$ with a single amplified pulse from a frequency-comb laser combined with HHG~\cite{Grundeman}. An excitation scheme was used based on one 790\,nm photon (near-infrared) and one 32\,nm photon (XUV), which is compatible with a future precision measurement based on RCS. Detection of $2S$ excitation in ~\cite{Grundeman} was performed with double ionization. For a precision measurement, a nondestructive readout technique such as QL is desirable, as it reduces the need for complicated and time-consuming reloading of the ion and initial-state preparation \cite{Grundeman,gardner_multi-photon_2019, loh_rempi_2012}. In this study we therefore combine the aforementioned excitation method with QL nondestructive state detection in an ion trap. It is based on a single trapped He$^+$ ion (spectroscopy ion) in a Paul trap together with a single laser-cooled Be$^+$ ion  (logic ion) for sympathetic cooling and nondestructive readout. We outline below the specific challenges associated with this endeavor, the methods. we propose to address them, and the aspects that are broadly relevant for combining XUV spectroscopy with quantum control techniques.

One challenge in combining QL with XUV light arises from the large recoil imparted on atoms upon absorption of XUV photons. In quantum logic spectroscopy (QLS)~\cite{brewer_Al_2019,King2022,hb3c-dk28,ma_quantum-logic-based_2024,dawel2026alclock16times1018systematic}, experiments are typically performed in the Lamb-Dicke regime, where the motional energy spacing of the trapped ion is larger than the recoil energy $E_\textrm{rec} = (\hbar k)^2 / 2m$, with $m$ the mass of the ion and $k=2\pi/\lambda$ the wavevector of the excitation light. In this regime, it is possible to perform recoil-free and (first-order) Doppler-free excitation to avoid motional shifts and motional broadening~\cite{dicke_effect_1953, wineland_experimental_1998}. However, reaching the Lamb-Dicke regime is difficult in the XUV range due to the short excitation wavelength. Although Doppler-free excitation with two identical wavelength, counter-propagating photons could in principle mitigate this~\cite{herrmann2009feasibility,moreno2023}, our excitation scheme based on co-propagating photons of different wavelengths significantly increases the excitation probability and eliminates the need for difficult to achieve spatial and temporal overlap of the counterpropagating pulses~\cite{Grundeman}. Therefore, we propose and numerically validate a method that circumvents both the recoil shift and the first-order Doppler effect by synchronizing the ion secular period to the time delay between the RCS laser pulses (the laser repetition period). Notably, this synchronization method is also applicable to single-photon XUV transitions, where Doppler-free excitation via two counter-propagating photons is not possible. The proof-of-principle has been demonstrated in the visible regime using a single trapped Ca$^+$ ion [S. Noel et al, in preparation].

Another challenge stems from the inherently low conversion efficiencies achievable in HHG, which for weak transitions such as the $1S-2S$ transition of He$^+$ leads to a low excitation probability and thus a low signal-to-noise ratio (SNR). QL readout is based on the ability to distinguish between two electronic states of the spectroscopy ion by interrogating a carefully prepared logic ion whose motional state depends only on the state of the spectroscopy ion. The logic ion features a motional state-dependent transition between dark and bright electronic states used to determine the spectroscopy ion's state (see \ref{QLS_sec}). This makes QL readout protocols very sensitive to false positive counts induced by the residual thermal energy of the logic ion through a non-zero population of excited motional states. This ultimately leads to a higher uncertainty on the measured transition frequency. To circumvent that, we propose a QL scheme based on higher-order red motional sideband readout, which enables detecting the motional excitation with high fidelity using Rapid 
Adiabatic Passage (RAP). We also present a numerical model for the motional dynamics of a trapped ion excited outside the Lamb-Dicke regime by two XUV pulses separated in time, as employed in RCS, and use it to study the impact of imperfect state preparation and quantum projection noise on QL readout. Our model can be easily adapted to consider an arbitrary number of excitation pulses, allowing for its application to direct frequency-comb spectroscopy on trapped ions.

Finally, we address an important ongoing challenge in the field of XUV spectroscopy, which is the phase coherence of the XUV light. While HHG is a phase-coherent process, the upconversion from the driving wavelength $\lambda$ to the $n^{\text{th}}$ harmonic $\lambda/n$ is accompanied by an increase of the phase noise power spectral density proportional to $n^2$. This can cause rapid decoherence of the light field, especially at very short wavelengths, thus limiting the Ramsey interrogation times and therefore the achievable precision. For intra-cavity HHG of frequency comb lasers, the cavity enhancement also reduces the phase noise on the OFC pulses by spectral filtering~\cite{moreno2023}. However, the filtering action and enhancement function are not independent parameters. For RCS, enhancement is performed by amplification, so this constraint is not present. We therefore propose an empty (vacuum) optical-filtering cavity for pure filtering purposes to extend the XUV coherence time, limited only by the dispersion management of the mirrors. The effect is simulated by modeling the phase noise of our OFC together with the filtering action of the cavity. The presented technique is universal and can be used in any experiment where coherent XUV light is required. We include the phase noise model in our simulations of the measurement of the $1S-2S$ transition frequency via RCS and QL readout, and conclude that a precision measurement at the 10$^{-12}$ fractional uncertainty level is realistic using the proposed methods.

The article is organized as follows. In Section~\ref{RCS_technique}, we give a brief overview of the RCS technique and discuss how it can be applied to measure the $1S-2S$ transition in trapped and sympathetically laser-cooled He$^{+}$, including a presentation of our QL readout scheme. In Section~\ref{RCS on trapped ions}, we provide a theoretical framework for numerical simulations of the motional excitation dynamics of trapped ions outside the Lamb-Dicke regime, excited with time-separated pulses. In Section~\ref{numerical_simu}, we provide numerical results on the interference of motional levels in Ramsey-type interrogation of trapped ions and present a technique to circumvent Doppler broadening and the recoil shift in trapped ions excited with co-propagating photons. We also discuss the use of higher-order red sideband RAP readout to minimize false positive counts induced by residual logic ion motional energy in our QL readout scheme. In Section~\ref{laser phase noise}, we present our phase noise model and simulation results. Finally, in Section~\ref{feasibility section}, we combine the motional dynamics and phase noise models to carry out Monte Carlo simulations of RCS of the $1S-2S$ transition in trapped He$^{+}$ using a QL readout scheme and double-ionization detection scheme. We provide estimates for the parameter ranges that would allow for a measurement with 10 kHz uncertainty.

\section{Ramsey-comb spectroscopy}\label{RCS_technique}
\begin{figure}
\includegraphics[width= \columnwidth]{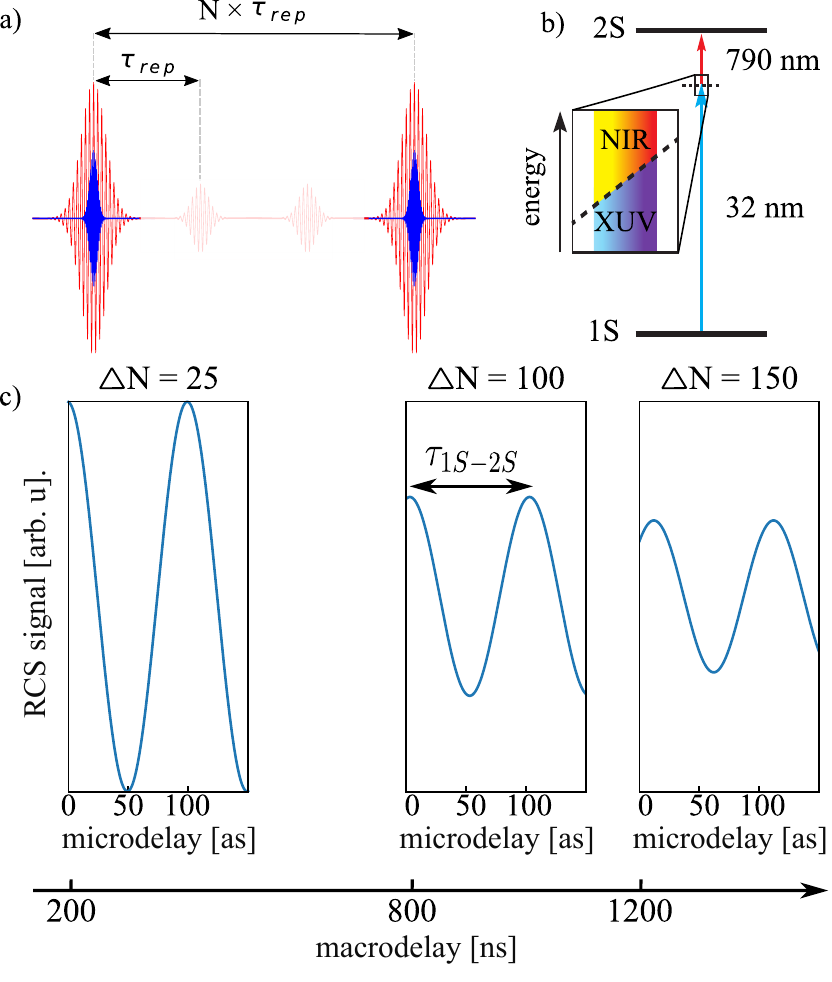}
\caption{\label{fig:RCS_scheme} Depiction of Ramsey-comb spectroscopy of the $1S-2S$ transition in He$^{+}$. a) Pairs of phase-coherent, two-color 790 nm NIR (red) + 32 nm XUV (blue) pulses derived from a FC and separated in time by multiples of the FC repetition period $\tau_{\textrm{rep}}$ are used to perform a series of individual Ramsey-type measurements. The 32 nm light is the 25$^{\textrm{th}}$ harmonic of the NIR fundamental and is generated by HHG in a gas jet. b) The 40.81 eV energy gap of the 2-photon $1S-2S$ transition in He$^{+}$ is bridged by a 790 nm photon and its 25$^{\textrm{th}}$ harmonic in the XUV at 32 nm. Inset: Different wavelength combinations within the bandwidths of the NIR and XUV pulses match the $1S-2S$ energy gap and their contributions coherently add up to enhance the transition probability. Note that this represents only the excitation process itself per pulse; the Ramsey fringes seen in the $2S$ excitation probability result from two such excitation events. c) A series of Ramsey fringes, showing the $2S$ population as a function of the delay between pairs of two-color, 790 nm + 32 nm excitation pulses. The graphs show the signal for three different macrodelays $\Delta N$, in units of the 8 ns RCS laser repetition period. As the macrodelay increases, RCS fringes lose contrast mainly due to an increase in the relative phase jitter between the two RCS excitation pulses, as discussed in section \ref{laser phase noise}.}
\end{figure}

Ramsey-comb spectroscopy (RCS) is a technique for high-precision spectroscopy of atomic and molecular systems \cite{Jonas, Altmann_H2, Charlaine, Laura_xenon}. It makes use of pairs of selectively amplified and phase-coherent ultrashort laser pulses derived from a frequency comb laser (FC) to perform a series of optical Ramsey-type measurements at different interpulse delays \cite{Jonas_theory}. Each pulse from the RCS pulse pair can be considered to induce a superposition of the ground and excited states of a 2-level system. The contributions from the first and second pulse have a relative phase determined by the time delay between the pulses $\Delta t$ and their relative optical phase shift $\Delta \phi$ (see eq. (\ref{RCS signal})). Interference between the two excited state contributions gives rise to a delay-dependent excited state population
\begin{equation}\label{RCS signal}
    |c_{e}(\Delta t)|^{2} \propto 1+\cos({2\pi f_{\textrm{tr}}\Delta t + \Delta \phi})
\end{equation}
where $f_{\textrm{tr}}$ is the transition frequency and where
\begin{equation}\label{delays}
    \Delta t = N\times\tau_{\textrm{rep}}
\end{equation}
with $N$ an integer and $\tau_{\textrm{rep}} = 1/f_{\textrm{rep}}$ the repetition period of the FC, which, for the RCS laser system, is 8 ns. $\Delta \phi$ represents the relative phase between the first and second RCS pulse and includes both the carrier-envelope phase shift intrinsic to frequency comb pulses, as well as any relative phase jitter between the pulses. 

We refer to the $\Delta t$ term as the interpulse delay and to $N$ as the macrodelay. $N$ is set by picking the second RCS pulse to be the $N^{\textrm{th}}$ pulse after the FC pulse chosen to be the first pulse. For a given macrodelay, $\Delta t$ can be adjusted by changing the FC repetition frequency $f_{\textrm{rep}}$ in order to sample the sinusoidal oscillation of the excited state population, typically over hundreds of attoseconds, from which $f_{\textrm{tr}}$ is determined \cite{Jonas, Altmann_H2, Charlaine}. For convenience, we can express the interpulse delay as
\begin{equation}\label{interpulse_delay_eq}
    \Delta t = N \times \tau_{\textrm{rep}} + \delta t
\end{equation}
and refer to the attosecond scale changes to $\Delta t$ given by $\delta t$ as the microdelay. Eq.(\ref{interpulse_delay_eq}) is an approximation because the small scale scanning is achieved by tuning the repetition frequency. It nevertheless provides a sufficiently accurate conceptual framework that is compatible with the numerical simulations presented in section \ref{numerical_simu}.

We refer to a measurement of the excited state population as a function of delay as a Ramsey fringe. RCS is based on combining a set of Ramsey fringes at various macrodelays into a single signal, thereby reducing some systematic effects and increasing the sensitivity of the fitting procedure used to determine the transition frequency (see Ref.~\cite{Jonas_theory}). The maximum macrodelay sets the frequency resolution of an RCS measurement and is typically limited by either laser phase noise or the excited state lifetime (and in the case of a beam experiment also the time the atoms or molecules spend in the interaction zone). In Ref.~\cite{Charlaine}, macrodelays as high as $N = 45$ (corresponding to 360 ns delay) were used at 201 nm, resulting in relative frequency uncertainty on the order of $1.3 \times 10^{-11}$.

Because RCS relies only on pairs of pulses, much higher peak powers can be reached than in full repetition rate FC amplification, without generating excessive thermal load. This makes RCS suitable for spectroscopy of weak transitions at short wavelengths for which frequency up-conversion processes are highly inefficient and where no continuous-wave laser sources are presently available. In addition to reaching into the XUV wavelength spectral region, RCS has the notable advantage of being insensitive to any delay-independent, common-mode effects \cite{Charlaine, Altmann_H2}.

\subsection{Ramsey-comb spectroscopy of the $1S-2S$ transition in He$^{+}$}
RCS of the $1S-2S$ transition in He$^{+}$ is schematically depicted in Fig. \ref{fig:RCS_scheme}. In this case, we make use of pairs of intense ($I \sim 10^{14}$ W/cm$^{2}$), time-separated and phase-coherent pulses derived from a FC. With these pulses we perform, as discussed before, a series of Ramsey-type measurements at increasing macrodelay. RCS of the 1S - 2S transition in He$^{+}$ will be performed with the same 2-color excitation scheme that we demonstrated in \cite{Grundeman}. As depicted in a), each of the two RCS pulses consists of a 790 nm NIR pulse and its 25$^{\textrm{th}}$ harmonic at 32 nm, produced by HHG as in Ref.~\cite{Laura_xenon}. Because we only use two pulses from the fundamental NIR frequency comb, the spectrum of the NIR and XUV radiation does not represent a narrow frequency comb spectrum as is seen for an infinite pulse train. The individual NIR pulses have a bandwidth of about 6 THz while the individual XUV pulses have a bandwidth of around 20 THz \cite{Grundeman}. Note that two such pulses with a fixed phase relation have a spectrum that resembles the single-pulse spectrum, but with a cosine modulation that has a period equal to the inverse of the pulse delay. As shown in the inset of Fig. {\ref{fig:RCS_scheme}} b), various wavelength combinations (frequency modes) within the NIR and XUV bandwidths match the transition energy, and their contributions add up coherently to enhance the transition probability. The frequency resolution of the measurement scheme is not determined by the bandwidth of the individual 2-color pulses but by the maximum interpulse delay used. The approximately 1.9 ms lifetime of the 2S state means that, in practice, RCS fringe contrast decay as a function of delay will not be lifetime-limited but rather due to delay-dependent noise sources, notably laser phase noise in the XUV as discussed in section \ref{laser phase noise}. The RCS laser system and vacuum setup are depicted in Fig. \ref{fig:experimental_setup} in the appendix. See Refs.~\cite{Grundeman, Charlaine} for further details.

RCS on He$^{+}$ requires confining the ion in an ion trap in order to interact with both RCS pulses (not possible in an atomic beam configuration due to the high velocity of He$^{+}$ and the small focus of the XUV beam). Ion trapping also allows for cooling techniques and first-order Doppler-free measurements with one-sided excitation. As He$^{+}$ does not have a suitable laser cooling transition, sympathetic cooling \cite{roth2005, Larson1986} with a co-trapped Beryllium ion can be used to cool the He$^{+}$ ion to its motional ground state \cite{sideband_cooling}. Initially, a broad ($2\pi\times 19.4$ MHz) Doppler cooling transition $2S_{1/2}\ket{F=2,m_{f}=2} \rightarrow 2P_{3/2}\ket{F=3,m_{f}=3}$ at 313 nm can be used to reach the $\sim 0.5$ mK Doppler cooling limit. Subsequently, the Be$^{+}$ ion can be cooled to the motional ground state by driving red sideband (RSB) stimulated Raman transitions between the $2S_{1/2}\ket{F=2,m_{f}=2}$ and the  $2S_{1/2}\ket{F=1,m_{f}=1}$ hyperfine sublevels, as initially demonstrated in \cite{sideband_cooling}. The Be$^{+}$ ion will serve not only for cooling but also as the logic ion in a quantum logic readout scheme as discussed in section \ref{QLS_sec}.


\subsection{Electronic excitation dynamics}\label{electronic_excitation_dynamics}
We perform time-dependent Schr\"odinger equation (TDSE) simulations of the two-color (790 nm + 32 nm) Ramsey-comb spectroscopy interrogation of He$^{+}$ to get insights into its electronic excitation dynamics \cite{Grundeman} of the system. We solve TDSE consisting of a 3D atomic (Coulomb) Hamiltonian and an atom-field interaction term (see eq. (\ref{electronic excitation hamiltonian}) in the appendix). The high harmonic field used in the simulations was generated by simulating the HHG process under the strong-field approximation for a single atom \cite{lewenstein, Jin} as in \cite{Grundeman}. For more details on the formalism and computational techniques used for these TDSE simulations, see section \ref{TDSE_simulations_appendix} in the appendix.

Fig. \ref{fig:fringes_fig} presents simulated Ramsey fringe data and sinusoidal fits for the excited 2S population (blue curve) and He$^{2+}$ (orange curve) ions produced with a NIR intensity of $4\times 10^{13}$ W/cm$^{2}$ and a 25$^{\textrm{th}}$ harmonic intensity of $2\times 10^{9}$ W/cm$^{2}$. The NIR and XUV intensities are an estimate of what we can reach with our laser system in the excitation region \cite{Grundeman}. In these simulations, He$^{2+}$ is produced mainly by the second RCS pulse through the 2S state, as was experimentally shown with a single pulse in \cite{Grundeman}. With these intensities, the 2S state reaches a maximum excitation probability slightly above $0.2\%$, while He$^{2+}$ reaches just above $0.04\%$. The lifetime of the relevant excited states other than the 2S is too short to have coherent interaction with both RCS excitation pulses. These states therefore do not produce Ramsey signals. The TDSE simulations also provide insights into the population dynamics during the excitation process itself, as shown in the appendix, section \ref{TDSE_simulations_appendix}. 
Fig.~\ref{fig:fringes_fig} will be used to make an initial estimate of the contrast loss produced by laser phase noise in the XUV, as presented in section \ref{laser phase noise}. 

\begin{figure}
\includegraphics[width=\columnwidth]{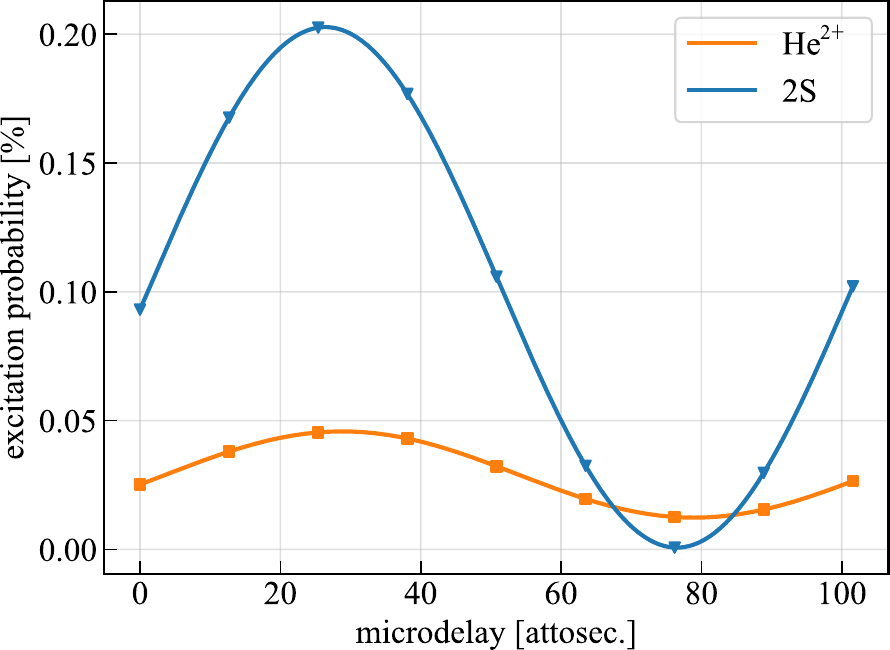}
\caption{\label{fig:fringes_fig} Simulated RCS fringes for $1S-2S$ excitation of He$^{+}$, for a free and static He$^{+}$ ion at a macrodelay of 600 fs (chosen for computational convenience). Solid lines are sinusoidal fits to the simulated data. For a NIR intensity of $4\times 10^{13}$ W/cm$^{2}$ we observe a sizable population of the 2S state (blue triangles) and He$^{2+}$ production due to 9-photon ionization of the 2S state (orange squares) at a maximum level of $\sim 0.2\%$ and $0.04 \%$ respectively.}
\end{figure}

\subsection{Quantum logic readout of $1S-2S$ excitation}\label{QLS_sec}
In prior RCS experiments, readout was performed using state-selective ionization of the excited state followed by ion extraction using an electric field and subsequent detection with an ion detector \cite{Charlaine, Laura_xenon, Altmann_H2}. RCS in an ion trap, however, opens up new readout possibilities, which come with their own set of challenges. Although state-selective ionization remains an option, the alternative, non-destructive quantum logic (QL) schemes are desirable as they avoid the need to reload an ion after each successful electronic excitation. Because the single-pulse $1S-2S$ transition probability is very low ($\sim 10^{-3}$ for our excitation scheme), a precision measurement requires many excitation attempts with a minimal ion loss/re-loading time.

\begin{figure*}
\includegraphics[width = 0.98\textwidth]{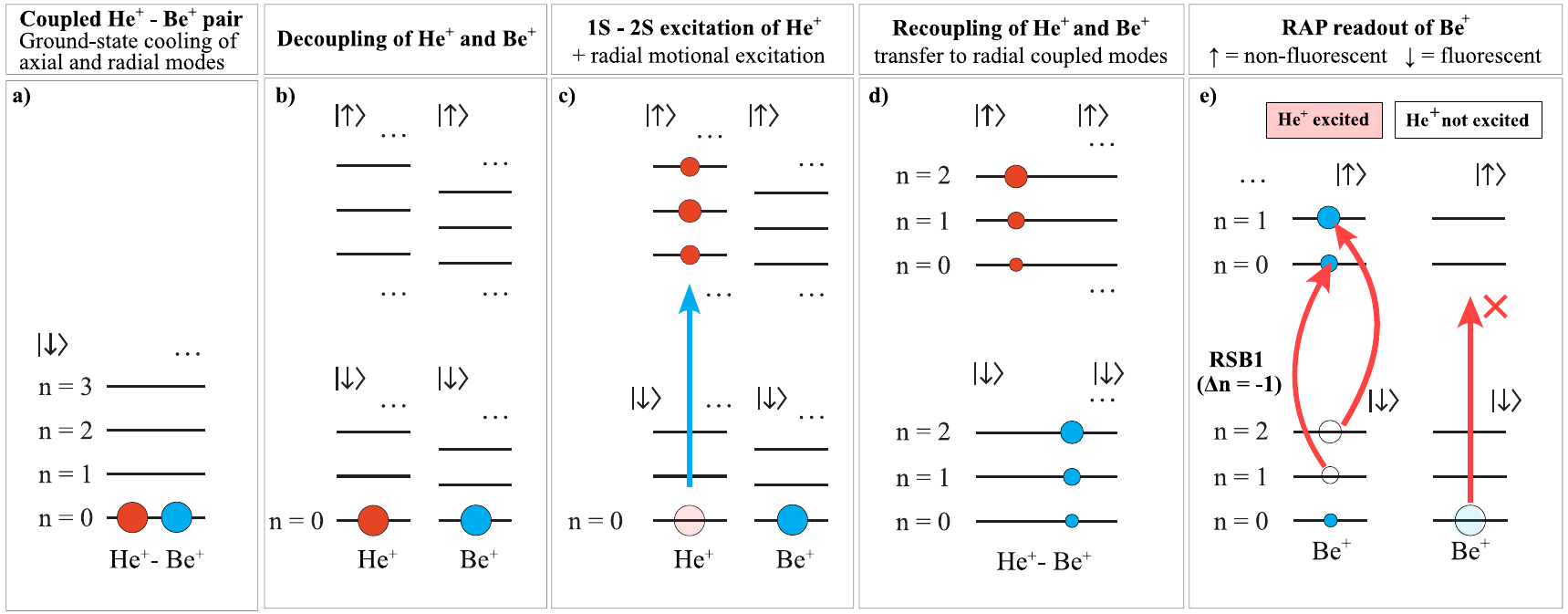}
\caption{\label{fig:quantum logic-type detection scheme} Quantum logic-type detection scheme based on single He$^{+}$ and Be$^{+}$ ions co-trapped in a linear Paul trap. For He$^{+}$, $\ket{\downarrow} = \ket{1S}$ and $\ket{\uparrow} = \ket{2S}$. For Be$^{+}$,  $\ket{\downarrow} = 2S_{1/2}\ket{F=2, m_{f}=2}$ and  $\ket{\uparrow} = 2S_{1/2}\ket{F=1, m_{f}=1}$. a) The coupled motional modes of He$^{+}$ with Be$^{+}$ are cooled to the ground state with both ions in their electronic ground state. b) The ions are radially motionally decoupled by relaxing the ion trap axial potential. c) Excitation of He$^{+}$ to the electronic excited state $\ket{\uparrow}$ by XUV radiation is accompanied by a large momentum transfer which promotes the He$^{+}$ ion to a coherent motional state with average motional quanta $\Bar{n} = 11$ distributed among its radial motional modes; The Be$^{+}$ remains in $\ket{\downarrow;n=0}$. Following RCS excitation, the axial potential is tightened, coupling the ions and distributing the energy in the He$^{+}$ modes to the shared modes. d) Readout is performed on Be$^{+}$ by RAP population transfer to the $\ket{\uparrow; n-o}$ state, where $o$ is the RSB order, followed by fluorescence detection. The population transfer is performed by addressing the Be$^{+}$ ion on the IP radial motional mode of the two-ion crystal.}
\end{figure*}

QL schemes have been implemented in atomic and molecular ions in the Lamb-Dicke regime \cite{wolf2016nondestructive, quantum_logic}. In this work, we propose a QL scheme in which the spectroscopy ion is far away from the Lamb-Dicke regime, as is the case for XUV spectroscopy of He$^{+}$ and other ions in which the imparted photon recoil energy exceeds the motional level separation. 
\subsubsection{Electronic and motional excitation of He$^{+}$}
Fig. \ref{fig:quantum logic-type detection scheme} outlines our proposed excitation and QL readout scheme. Following sympathetic Doppler cooling, the coupled motional modes of the He$^{+}$- Be$^{+}$ crystal are cooled to the motional ground state by Raman sideband cooling \cite{sideband_cooling}, as shown in Fig. \ref{fig:quantum logic-type detection scheme} a). The ions are then spatially separated axially and motionally decoupled along the radial direction by relaxing the axial potential of the ion trap. This is done to avoid double-ionization of Be$^{+}$ from the XUV radiation and for first-order Doppler effect suppression as discussed in section \ref{doppler_suppression} and shown in Fig. \ref{fig:quantum logic-type detection scheme} b). With an axial separation of 40 micrometers between the ions, we calculate the contribution of Be$^{+}$ to the mode amplitude of the radial center-of-mass mode at 5 MHz to be approximately 5$\times 10^{-4}$ for our linear Paul trap, effectively radially decoupling the ions. As explained in detail in section \ref{doppler_suppression}, first-order Doppler suppression is implemented by synchronizing the radial secular frequency of the radially decoupled helium ion to the FC repetition period. Following radial decoupling, electronic excitation on the spectroscopy (He$^{+}$) ion along the radial direction (normal to the page) is attempted (Fig. \ref{fig:quantum logic-type detection scheme} c)). If He$^{+}$ was successfully excited to the 2S electronic excited state, it receives a large momentum transfer from the co-propagating NIR+XUV photons, displacing the ground state and promoting it to a coherent motional state with average motional occupation number $\Bar{n}=11$ of the radial motional mode, similar to the situation in photon recoil spectroscopy \cite{Wan2014PhotonRecoil}.

\subsubsection{QL readout of He$^{+}$ excitation}
From the moment of separation, the Be$^{+}$ logic ion can continue to be cooled on both the axial modes (which are still strongly coupled with He$^{+}$) and radial modes to keep it close to the ground state. After attempting electronic excitation of He$^{+}$, the radial modes of the ions are coupled again by tightening the ion trap's axial potential, as shown in Fig. \ref{fig:quantum logic-type detection scheme} d). The He$^{+}$ motional energy is transferred to the radial coupled modes of the He$^{+}$- Be$^{+}$ Coulomb crystal. At a separation of approximately 8 $\mu$m, the mode vector amplitudes for the in-phase (IP) mode are 0.974 and 0.226 for He$^{+}$ and Be$^{+}$, respectively, allowing for high readout transfer efficiency using RAP as explained below. Smaller inter-ion separation risks pushing the crystal into a radial orientation.

Readout proceeds in two steps. First, population transfer is attempted using RSB Rapid Adiabatic Passage (RAP) \cite{Gebert_2016, Gebert_2018_corrigendum} from the Be$^{+}$ electronic ground state, lower hyperfine level $\ket{\downarrow} = 2S_{1/2}\ket{F=2, m_{f}=2}$ to the electronic ground state, upper hyperfine level $\ket{\uparrow} = 2S_{1/2}\ket{F=1, m_{f}=1}$. The participation of He$^{+}$ in the IP coupled radial mode is larger than in the out-of-phase mode. Population transfer is therefore implemented by addressing the IP mode. This population transfer is only possible if He$^{+}$ was excited to the 2S state and higher motional states ($n > 0$) of the shared He$^{+}$ - Be$^{+}$ radial modes were populated. After attempting population transfer from  $\ket{\downarrow;n}$ to $\ket{\uparrow;n-o}$, with $o$ the RSB order ($o= 1,2,3,...$), the Be$^{+}$ is addressed with the 313 nm Doppler cooling beam. If Be$^{+}$ remains in $\ket{\downarrow}$, fluorescence is detected on the Doppler cooling transition and it appears bright, while the ion remains dark if Be$^{+}$ was transferred to the $\ket{\uparrow}$. These steps are shown in Fig. \ref{fig:quantum logic-type detection scheme} e).

\begin{figure}[h!]
\includegraphics[width=0.85 \columnwidth]{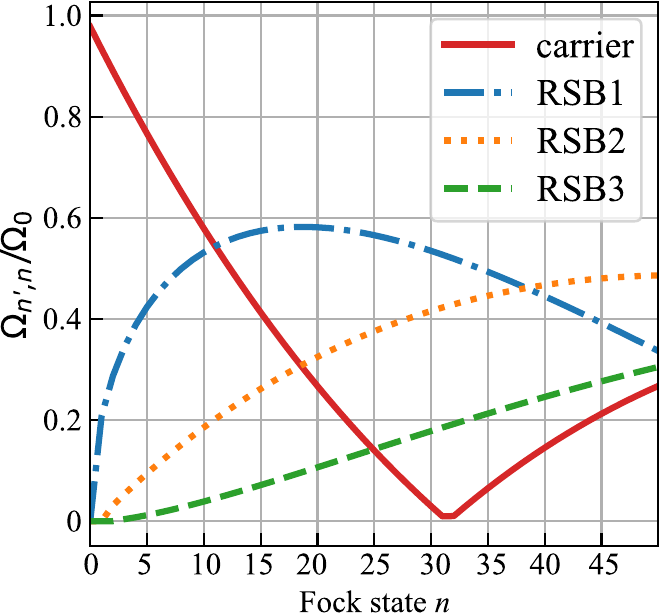}\caption{\label{fig:Rabi_freq_fig}
Normalized Rabi frequency for carrier (solid red), first- (dash-dotted blue), second- (dotted orange), and third-order (dashed purple) red sidebands for a Lamb-Dicke parameter of $ \eta = 0.2$.}
\end{figure}

RAP allows for population transfer between bare atomic levels which is robust against fluctuations in experimental parameters such as laser intensity and frequency. Furthermore, it allows for high transfer efficiencies across a range of coupling strengths (Rabi frequencies) \cite{Noel_2012}. This makes the technique well suited to carry out population transfer on trapped ions with sizable probability amplitudes for a range of Fock states, each of which has a different Rabi frequency. Fig. \ref{fig:Rabi_freq_fig} shows how Rabi frequencies $\Omega_{n^{\prime},n} = \Omega_{0} C_{n^{\prime},n}$ differ not only between motional levels, but also between sideband orders, with $\Omega_{0}$ the bare Rabi frequency and with $C_{n^{\prime},n}$ the matrix elements of the displacement operator for $n^{\prime} \rightarrow n$ motional transitions (see appendix). For a Be$^{+}$ mode amplitude of 0.226, we expect $\Omega_{0} = 2\pi \times 16$ kHz on the IP radial mode, leading to expected transfer efficiencies of at least $90\%$ \cite{Gebert_2016}.

This detection technique suffers from false positive counts due to thermal occupation of higher motional levels and imperfect readout of the Be$^{+}$ ion. In section \ref{Temperature and red sideband order readout} we use numerical simulations of the RCS excitation process to assess the impact of such false positive counts and show how it can be reduced by using higher-order RSB transitions to the $2S_{1/2}\ket{F=1,m_{f}=1}$ state together with RAP. The numerical RCS excitation model is developed in the following section (\ref{RCS on trapped ions}). 

\section{Theory of the motional excitation dynamics of trapped ions excited by time-separated pulses}\label{RCS on trapped ions}

We consider a He$^{+}$ ion which is motionally decoupled in the radial direction from the Be$^{+}$ ion and assume a $1S-2S$ transition probability of He$^{+}$ of approximately $10^{-3}$ per single NIR+XUV pulse. This is based on TDSE simulations of the electronic excitation process (see \ref{electronic_excitation_dynamics}). We only consider the radial motional mode  as that is the direction of the excitation laser. 


We describe the initial motional state of the ion by a linear superposition of Fock states with random phases and with probability amplitudes corresponding to a Boltzman distribution for an ion in thermal equilibrium with a reservoir at temperature T. This method allows us to incorporate the effects of nonzero coherences between the motional levels after interaction with the two RCS excitation pulses, and to include a random initial phase for each motional quantum state to replicate the properties of a thermal state for our Monte Carlo simulation. We therefore describe the initial motional state of the ion as
\begin{equation}\label{thermal dist}
     \ket{\beta} = \sum_{n=0}^{\infty} \beta_{n}\ket{n}
\end{equation}
characterized by the complex numbers $\beta_{n} = B_{n}e^{i\phi_{n}}$, with $B_{n}$ the weights (amplitude) and $\phi_{n}$ the initial phases of each motional level. The weights are proportional to the square root of the probability to find the ion in a given motional state $B_{n} = \sqrt{P_{n}}$
with the probability $P_{n}$ given by \cite{Leibfried}
\begin{equation}\label{level prob}
    P_{n} = \frac{\Bar{n}^{n}}{(\Bar{n}+1)^{n+1}}
\end{equation}
where the mean occupation number $\Bar{n}$ is given by
\begin{equation}\label{Boltzmann factor}
    \Bar{n} = \frac{1}{\left(e^{\hbar\omega_{\textrm{sec}}/k_{B}T}-1 \right)}
\end{equation}
with $k_{B}$ Boltzmann's constant and $\omega_{\textrm{sec}}$ the ion secular frequency. 
 
\begin{figure}
    \centering
    \includegraphics[width=\columnwidth]{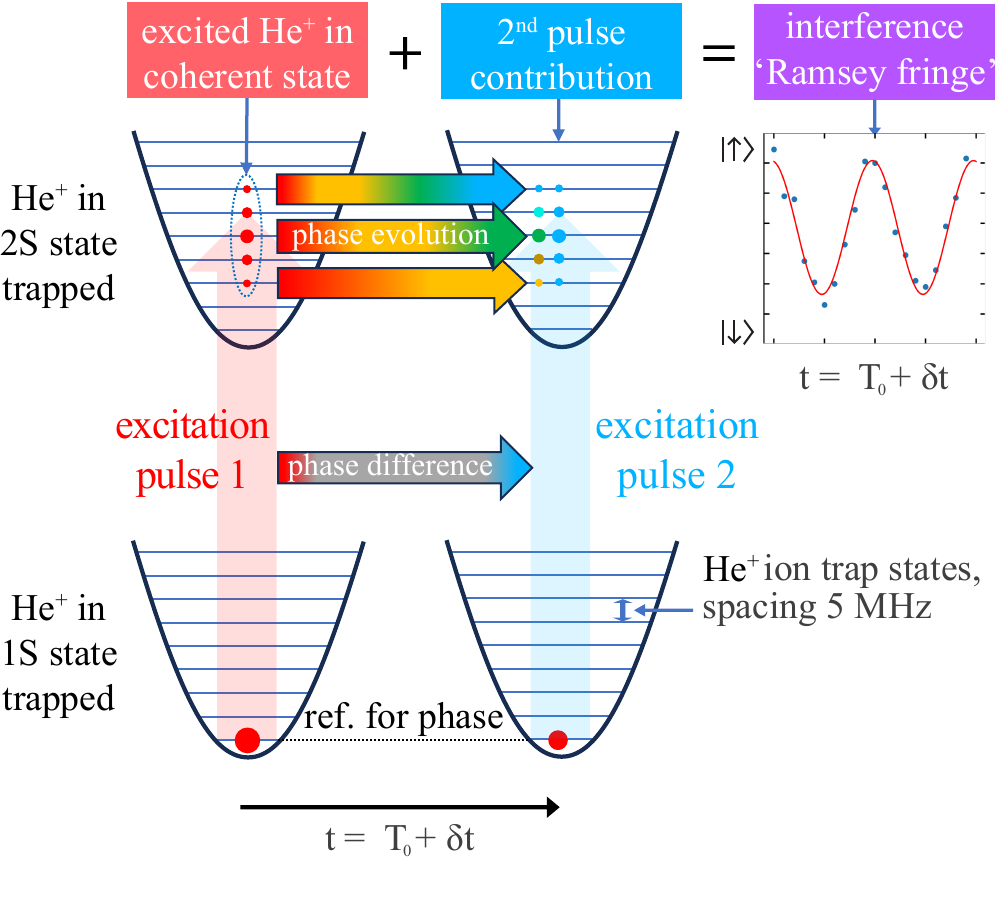}
    \caption{Schematic depiction of RCS of He$^{+}$ in an ion trap. The He$^{+}$ ion is decoupled from the Be$^{+}$ ion and initially in the motional and electronic ground state (see part b of Fig.~\ref{fig:quantum logic-type detection scheme}). The first RCS pulse, depicted as the light red vertical arrow, creates a superposition of the ground electronic and excited state. The electronic excited state features several occupied motional levels. The phase of the excited state wavefunction evolves as $\phi_{k}(t) = \omega_{tr} t + \omega_{sec,k}t$. Each motional level evolves at its own frequency $\omega_{sec,k} = k\times\omega_{\textrm{sec}}$, where $\omega_{\textrm{sec}}$ is the ion secular frequency and $k$ indexes the motional level. At a later time $t$ the second excitation pulse, depicted as the light blue vertical arrow, interacts only with the ground state part of the wave function (to leading order) and creates a second superposition of ground and excited motional and electronic states. The two contributions to the electronic excited state interfere constructively or destructively depending on their relative phases.}
    \label{fig:RCS_in_trap_overview}
\end{figure}

We start with an ion wavefunction, at time $t=0$ and temperature T, given by
\begin{equation}\label{ion_starting_wavefunction}
    \psi(t=0)=\ket{\beta} \otimes \ket{\downarrow}
\end{equation}
where $\ket{\downarrow}$ represent the 1S electronic ground state and $\ket{\uparrow}$ the 2S electronic excited state. In order to derive the ion wavefunction after the two pulse RCS sequence, we consider an operator representing the interaction of the ion with a resonant laser pulse given by
\begin{equation}\label{RCS_operator}
    \hat{\mathcal O } = \lambda_{2}\hat{D}(\alpha)\otimes \hat{T}+\lambda_{1}\mathds{1}
\end{equation}
Because the details of the electronic excitation (given in section \ref{electronic_excitation_dynamics}) are not relevant for the effects considered in this section, we model the electronic excitation with the simple $1S-2S$ flip operator $\hat{T}$, acting on electronic states as $\hat{T}\ket{\downarrow}=\ket{\uparrow}$ and $\hat{T}\ket{\uparrow}=\ket{\downarrow}$, akin to the spin flip operator $S_{+}$ defined in terms of the Cartesian Pauli matrices \cite{Leibfried}. $\lambda_{2}$ is the excitation amplitude, taken as the square root of the $10^{-3}$ $1S-2S$ electronic excitation probability, $\lambda_{1}$ is the square root of the probability $P_{g} = \sqrt{1-\lambda_{2}^{2}}$ to remain in the electronic ground state, and $\hat{D}(\alpha)$ is the displacement operator for a displacement of amplitude $\alpha =\eta e^{i\xi}$ in complex phase space representing the recoil imparted onto the He$^{+}$ ion by the 790 nm and 32 nm photons. 

The Lamb-Dicke parameter $\eta$ for He$^{+}$ is given by $\eta = k\times x_{0}$, where $k$ is the total wave vector of the 790 nm and 32 nm photons and $x_{0} = \sqrt{\hbar/2m_{\textrm{He}}\omega_{\textrm{sec}}}$ is the spatial extent of the ground state wave function with $m_{\textrm{He}}$ the mass of the He$^{+}$ ion, and $\omega_{\textrm{sec}}$ is the radial secular frequency of the ion in the trap. For our simulations we take $\xi = \pi/2$, and $\omega_{\textrm{sec}} = 2\pi \times 5$ MHz, leading to $\alpha = i 3.3422$ which is purely imaginary, representing a pure momentum displacement, in contrast to a purely real $\alpha$ representing a pure spatial displacement, equivalent to a displacement of the ion trap potential (see \cite{Ziesel_2013} for an experimental realization of displaced number states by means of such a trap displacement). Lastly, $\mathds{1}$ is the identity matrix on the vector space of the tensor product state in eq. (\ref{ion_starting_wavefunction}). 

We obtain the ion wavefunction after the two-pulse RCS sequence by making use of the RCS operator eq. (\ref{RCS_operator}) for pulse 1 (P1) and pulse 2 (P2) and the time evolution operator $\hat{U}(t)$ as 
\begin{equation}\label{ion_wave_function_after_RCS}
   \ket{\psi(t^{\prime})_{RCS}} = \hat{\mathcal O }_{P2}\hat{U}(t^{\prime})\hat{\mathcal O }_{P1}\ket{\psi(t=0)}
\end{equation}
The time evolution operator for evolution from $t = 0$ to $t = t^{\prime}$  is given by
\begin{equation}\label{time_evolution_op}
    \hat{U}(t^{\prime}) = e^{-i\hat{H}t^{\prime}/\hbar}
\end{equation}
with $\hat{H} = \hat{H}_{m} + \hat{H}_{e}$, where $\hat{H}_{m} = \hbar \omega_{\textrm{sec}}(\hat{N}+1/2)$ is the harmonic oscillator motional Hamiltonian for a trapped ion with secular frequency $\omega_{\textrm{sec}}$, $\hat{N}$ being the number operator. $\hat{H}_{e}$ is the electronic Hamiltonian (see appendix, section \ref{TDSE_simulations_appendix}) which we take to act on the electronic wave functions as $\hat{H}_{e}\ket{\downarrow}=0$ and $\hat{H}_{e}\ket{\uparrow} = \hbar \omega_{tr} \ket{\uparrow}$, where $\omega_{tr}$ is the (angular) $1S\rightarrow2S$ transition frequency.

Fig. \ref{fig:RCS_in_trap_overview} depicts the process described by eq. (\ref{ion_wave_function_after_RCS}). The initial ion wave function in the electronic ground state is promoted to a superposition of the electronic ground and excited state. The electronic ground state contribution remains in the motional ground state while the electronic excited state contribution is promoted to a coherent motional state by the first RCS pulse, modeled by $\hat{\mathcal O }_{P1}$ and depicted as the faint red vertical arrow. The occupied motional levels of the electronic excited states feature a phase evolution according to $\hat{U}(t^{\prime})$, represented by the upper three horizontal arrows with varying color gradient steepness to represent their different phase evolution (determined by their energy). To leading order, the second RCS pulse, modeled by $\hat{\mathcal O }_{P2}$ and depicted as the faint light blue vertical arrow, only interacts with the part of the initial ion wavefunction which remained in the electronic and motional ground state (we ignore stimulated transitions back to the ground state or to higher excited states). It produces a second contribution to the electronic excited state. The two contributions to the excited state feature a relative phase, as depicted in the color differences between the phase-evolved motional states produced by pulse one (yellow, brown, green, cyan and blue circles), and those produced by pulse two (purely blue circles). Such additional motional phase evolution mimics a Doppler broadening as discussed in \ref{doppler_suppression}.  

Projecting the wavefunction given by eq. (\ref{ion_wave_function_after_RCS}) onto the 2S state in an arbitrary motional state $m$ yields (see appendix, section \ref{appendix_ion_derivation} for a full derivation)
\begin{multline}\label{RCS_prob_amp}
 \bra{\psi_{2S}}\ket{\psi(t^{\prime})_{RCS}} = \\
 \lambda_{2}\sum_{n,m}\beta_{n}C_{m,n}\times 
 \\ \left (e^{-i(\omega_{sec, n}t^{\prime}-\Delta\phi(t^{\prime}))} + e^{-i(\omega_{sec, m}+\omega_{tr})t^{\prime}}\right)
\end{multline}
where $C_{m,n}$ are matrix elements of the displacement operator (see Appendix, section \ref{appendix_ion_derivation}) and where $\omega_{sec,n} = n\times\omega_{\textrm{sec}}$ and $\omega_{sec,m} = m\times\omega_{\textrm{sec}}$ with $n$ the initial motional quantum number and $m$ the final motional quantum number. The relative phase term $\Delta \phi(t)$ is given by $\Delta \phi(t) = \phi_{CEO} + \delta \xi(t)$, with $\phi_{CEO}$ the carrier-envelope offset frequency intrinsic to frequency combs, and $\delta \xi(t)$ a relative phase noise term. The RCS signal is then given by
\begin{equation}\label{RCS_signal}
RCS_{\textrm{signal}} = |\bra{\psi_{2S}}\ket{\psi(t)_{RCS}}|^{2}
\end{equation}
and results in a Ramsey fringe as shown in the far right of Fig. (\ref{fig:RCS_in_trap_overview}) when sampling the interpulse delay over the transition period (=$1/f_{1S-2S}$). A detailed discussion of the achievable signal-to-noise ratio (SNR) is provided in section \ref{numerical_simu}.

\section{Suppression of the first-order Doppler effect and recoil shift in trapped ion spectroscopy}\label{doppler_suppression}

As demonstrated below, by synchronizing the RCS macrodelay (or repetition period) to an integer multiple of the ion secular period, the influence of the harmonic potential on the acquired phase drops out, leaving only the phase due to precession at the transition frequency $\omega_{tr}$, the initial random phase $e^{i \phi_{n}}$ (hidden in the $\beta_{n}$ term and effectively leading to noise for a thermal state), and the relative phase term $\Delta \phi(t)$. Inspecting equation (\ref{RCS_prob_amp}), we can see that for 
\begin{equation}\label{dopper_condition}
t^{\prime}=j\times 2\pi/\omega_{\textrm{sec}} + \delta t= j\tau_{\textrm{sec}}+\delta t 
\end{equation} with $j$ an integer, $\tau_{\textrm{sec}}$ the secular period, and $\delta t$ the RCS microdelay, the probability amplitude of the 2S state reduces to
\begin{equation}\label{RCS_prob_amp_sync}
 \bra{\psi_{2S}}\ket{\psi(t^{\prime})_{RCS}} =
 \lambda_{2}\sum_{n,m}\beta_{n}C_{m,n}\times \left (e^{i\Delta\phi(t^{\prime}))} + e^{-i\omega_{tr} t^{\prime}}\right)
\end{equation}
where we take the (at most) $10^{-9}$ contribution of $\omega_{\textrm{sec}} \times \delta t$ to be zero. In other words, because the microdelay range (on the order of 100 attoseconds) is so much smaller than the secular period (200 ns), we can take the Doppler-free condition eq. (\ref{dopper_condition}) to instead be given by $t^{\prime}=j\times\tau_{\textrm{sec}}$. 

Without such synchronization, there is a sharp loss of fringe contrast due to interference between the phases of the different motional levels. This phenomenon is depicted in Fig. \ref{fig:macrodelay_fig} and may be interpreted as a situation in which two so-called Schr\"odinger cat states (see e.g. Ref.~\cite{schrodinger_cat}), each created (up to a relative phase) from the initial state described by equation (\ref{ion_starting_wavefunction}) by a single RCS pulse, have no spatial overlap at a given time delay, thereby producing no RCS signal. For a radial secular frequency of $2\pi \times 5$ MHz, the spatial extent of the ground state ion wavefunction is $x_{0} = 15.9$ nm so that it only takes 4.8 ns for the cat state created by the first RCS pulse to have moved beyond the spatial extent of the cat state created by the second RCS pulse due to the recoil imparted by the 32 nm and 790 nm photons onto the He$^{+}$. Only when the two cat states are allowed to evolve in time to the point where they spatially overlap again, in other words when the ensemble of phases $\phi_{n}$ of the two cat states are in phase again, does the RCS signal reappear. 

This dephasing of motional levels and the subsequent fringe contrast loss is a quantum mechanical equivalent of classical Doppler broadening, viewed in the time domain. Synchronization of the macrodelay to the secular period is therefore an effective technique to suppress Doppler broadening in Ramsey-type measurements on trapped ions outside the Lamb-Dicke regime with single-sided excitation. In addition to Doppler-broadening, however, one has to consider the effect of the large recoil shift which results from excitation with co-propagating 790 nm and 32 nm photons.

Excitation with a 790 nm and a 32 nm photon induces a recoil shift of $f_{rec} = $ 54 MHz on $^{4}$He$^{+}$. The recoil promotes an initially motional ground-state cooled ion to a displaced coherent state with a Poissonian distribution of Fock states centered on $n = f_{rec}/f_{\textrm{sec}} \approx 11$, for a radial secular frequency of $f_{\textrm{sec}} = 5$ MHz. For perfect synchronization, $N \times \tau_{\textrm{rep}} = j \times \tau_{\textrm{sec}}$ (or $N/f_{\textrm{rep}} = j/f_{\textrm{sec}}$) with $j$ an integer, there is no relative phase shift between the excited state wave packet created by the first RCS pulse and that created by the second RCS pulse (see Fig. \ref{fig:RCS_in_trap_overview}). However, for a fractional deviation in synchronization of $\delta = (f_{\textrm{sec}} - f_{\textrm{rep}})/f_{\textrm{sec}}$, the excited state wave packet created by the first RCS pulse will acquire a motional phase shift relative to the excited state wave packet created by the second RCS pulse which increases linearly with delay, leading to a frequency shift of $\delta\times n f_{\textrm{sec}} = \delta \times f_{rec} = \delta \times $54 MHz. Synchronization of the ion secular period to the excitation laser repetition period therefore suppresses both the delay-independent first-order Doppler broadening and the delay-dependent recoil frequency shift, despite the fact that we use a single-sided excitation scheme. Ultimately, the precision of the synchronization accuracy $\delta$ combined with the magnitude of the recoil determines the accuracy of the spectroscopy. As $\omega_{\textrm{sec}}$ control has been shown at a level of $10^{-5}$~\cite{Johnson2016}, other sources of noise aside, a spectroscopic accuracy below one kHz seems ultimately feasible for the $1S-2S$ transition of He$^{+}$. 

Note that the proposed synchronization technique can be applied with any single-sided excitation scheme in the resolved sideband regime, regardless of the number of photons involved in the process, as long as the transition linewidth is smaller than the ion trap harmonic oscillator spacing $\hbar \omega_{\textrm{sec}}$. If that is not the case then spectroscopy leads to a distorted lineshape and associated systematic shifts, even in the Lamb-Dicke regime \cite{Wan2014PhotonRecoil, Schulte2018}.

\section{Numerical simulations of RCS on trapped ions}\label{numerical_simu}
Based on the formalism presented in section \ref{RCS on trapped ions}, we simulate the motional excitation dynamics of He$^{+}$ interrogated with the RCS scheme in the XUV. In this section, we start by showing results produced without any sources of noise to illustrate the purely quantum nature of the dephasing of motional levels for $N\times\tau_{\textrm{rep}} \neq  j \times \tau_{\textrm{sec}}$, where $\tau_{\textrm{sec}} = 200$ ns is the ion secular period in the radial direction. Then, in section \ref{Temperature and red sideband order readout} we incorporate quantum projection noise and readout noise due to residual ion thermal energy to investigate QL readout of weak transitions in trapped ions. 


\subsection{Matching of ion secular period and laser repetition period}\label{freq_matching}

\begin{figure}
\includegraphics[width = 0.9\columnwidth]{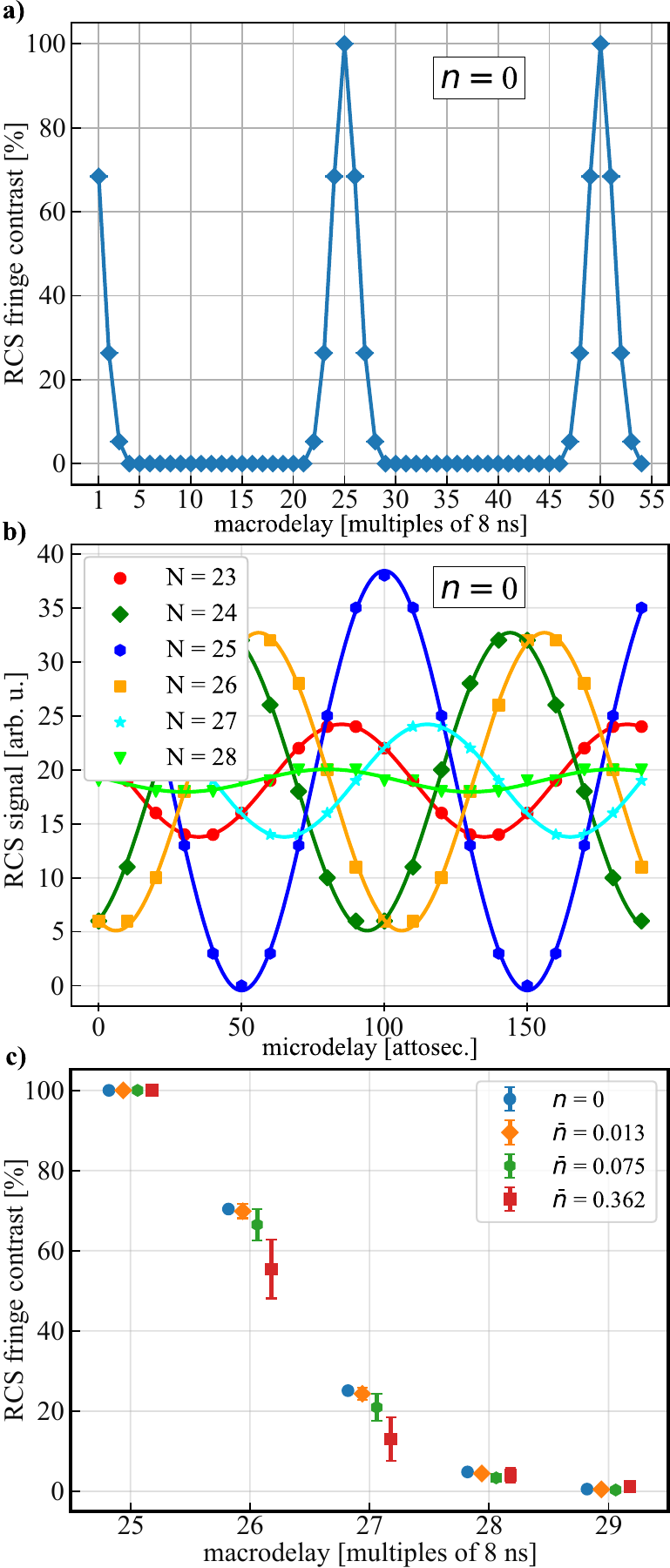}
\caption{\label{fig:macrodelay_fig} a) RCS fringe contrast as a function of macrodelay, in units of 8 ns, with linear interpolation between the data points to guide the eye. b) RCS fringes at various macrodelays, in units of 8 ns, corresponding to the data in a). c) average contrast as a function of macrodelay (note the artificial horizontal offsets for visibility) for four different initial He$^{+}$ ion average motional number $\Bar{n}$. Based on 100 simulations; the error bars shows the standard deviation of the average contrast per macrodelay. The temperature effect (reflected in $\Bar{n}$) is minor compared to dephasing of the induced coherent motional state. For perfect matching of the ion wavepacket motion and pulse repetition time there is no contrast loss from higher temperature, see section \ref{freq_matching}.}
\end{figure}


As explained in section \ref{doppler_suppression}, matching of the secular frequency of the ion with the RCS pulse macrodelay suppresses first-order Doppler broadening and the recoil shift. Here we investigate how these effects affect RCS fringe contrast. Fig. \ref{fig:macrodelay_fig} a) shows RCS fringe contrast, while \ref{fig:macrodelay_fig} b) shows the fringes it is calculated from. This is shown for macrodelays of 1 to 55, in units of the 8 ns RCS laser repetition period $\tau_{\textrm{rep}}$. These results are for an ion purely in the motional ground state ($n=0$) with a secular period $\tau_{\textrm{sec}} = 25 \times \tau_{\textrm{rep}} =$ 200 ns in the absence of all sources of noise. In Fig. \ref{fig:macrodelay_fig} a) the contrast loss due to de-phasing of the excited motional levels as a function of interpulse delay is clearly visible, as is the re-phasing at delays equal to an integer multiple of the ion's secular period (described by equations (\ref{RCS_prob_amp}) and (\ref{RCS_prob_amp_sync})).
Synchronization within $\Delta N = j\times 25 \pm 2$ with j an integer, is necessary to keep the contrast above 25\%, which is the practical minimum for analyzing RCS fringes. Fig. \ref{fig:macrodelay_fig} c) shows the average RCS fringe contrast as a function of macrodelay for 100 simulations, including noise due to the random initial phases $e^{i\phi_{n}}$ of motional levels $n$ (see eq. (\ref{thermal dist})), for a He$^{+}$ ion with initial average motional state occupation numbers of $\Bar{n} = $ 0, 0.013, 0.075 and 0.362. At higher temperatures and in the absence of any sources of noise, interference of higher occupied motional levels leads to a loss of contrast which is only slightly faster than that of a perfectly cold ion. When incorporating experimental noise, however, false positive counts due to residual ion population in higher motional levels coupled to this dephasing of motional levels leads to a faster drop in contrast, significantly more so when laser phase noise is taken into account. It therefore remains necessary to work at the best achievable synchronization, as is demonstrated in the following sections. 

\subsection{Impact of residual temperature and red sideband order on QL readout}\label{Temperature and red sideband order readout}

\begin{figure}[h!]
\includegraphics[width = \columnwidth]{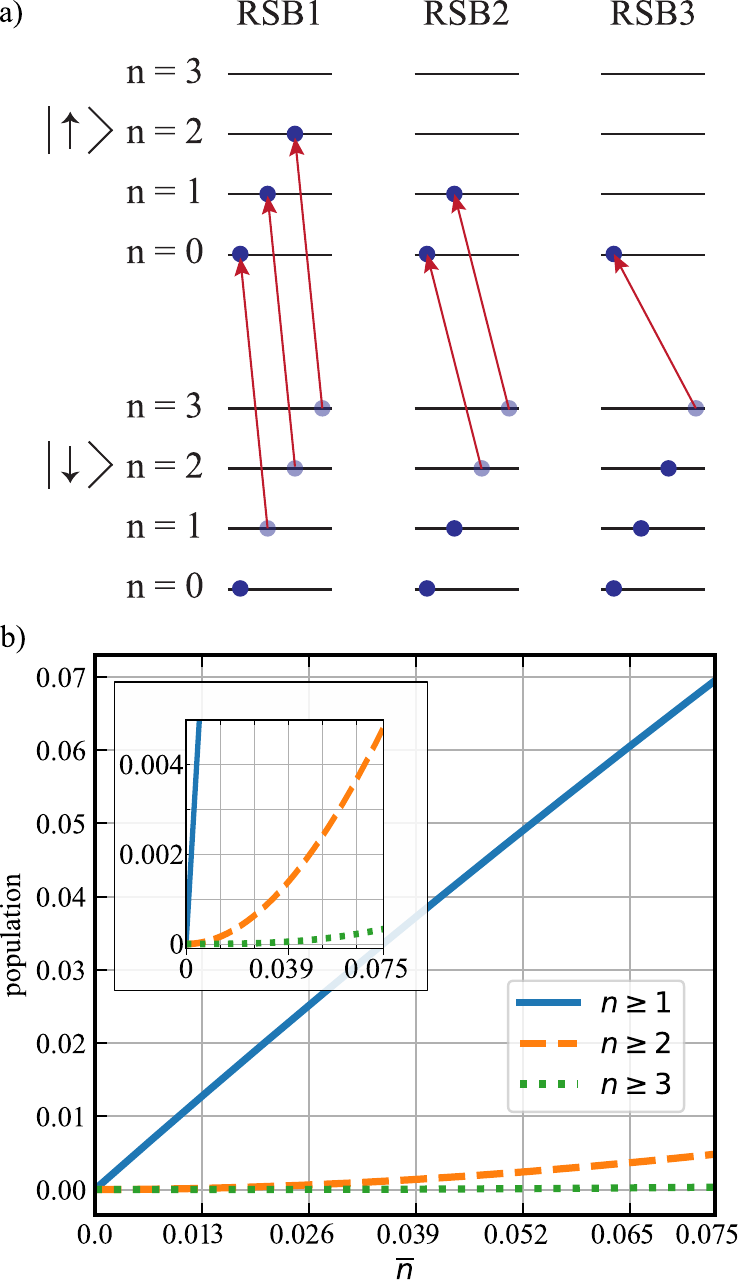}
\caption{\label{fig:RSB_order} Contribution to false positive counts due to higher occupied motional levels of the logic (Be$^{+}$) ion under different RSB order readouts using the logic scheme depicted in Fig. \ref{fig:quantum logic-type detection scheme}. a) Thermal occupation of motional levels at or above the RSB order can produce false positive counts. For clarity, only the first four motional levels are depicted. b) Motional level occupation probability for levels that can contribute to a false positive count as depicted in a) as a function of ion average occupation number. Integrated occupation probability for levels at $n\geq 1$ (RSB1) in solid blue, $n\geq 2$ (RSB2) in dashed orange, and $n\geq 3$ (RSB3) in dotted green. The inset is a zoom-in up to 0.005 population to show the RSB2 and RSB3 curves in more detail.}
\end{figure}

Occupation of excited motional modes ($n > 0$) due to imperfect state preparation and trap-induced ion heating can lead to false positives in our proposed QL readout scheme. As depicted schematically in Fig. \ref{fig:RSB_order} a), any population in motional levels at or above the order of the RSB readout pulse can contribute to such a false readout. Fig. \ref{fig:RSB_order} b) shows the integrated occupation probability of motional states at or above a given sideband order as a function of the average occupation number $\Bar{n}$. In our QL readout configuration, these correspond to radial motional states. The higher the $\Bar{n}$, the greater the probability of populating states that could lead to a false positive readout. The use of higher-order RSB RAP population transfer leads to suppression of false positives.

\begin{figure}[h!]
\includegraphics[width=0.675 \columnwidth]{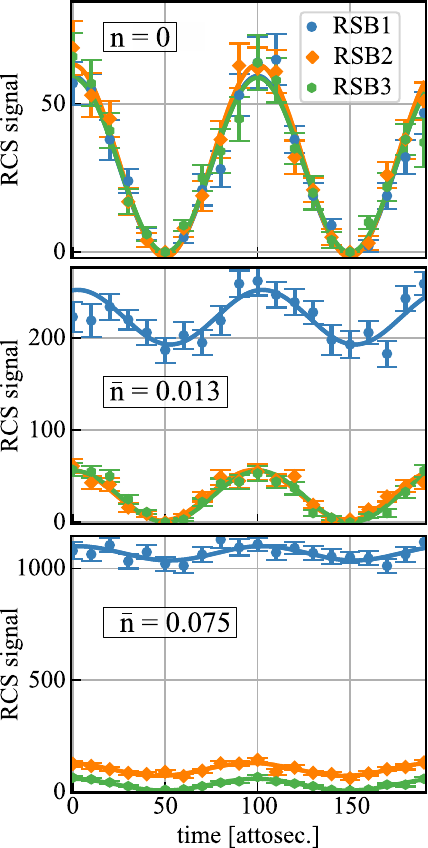}\caption{\label{fig:RSB_order_fringes}
RCS fringes read out with 1$^{\textrm{st}}$, 2$^{\textrm{nd}}$ and 3$^{\textrm{rd}}$ order RSB pulses on the Be$^{+}$ logic ion, for Be$^{+}$ average initial thermal occupation numbers of $\Bar{n} =$ 0, 0.013, and 0.075 at macrodelay $\Delta N$ = 25. Note the difference in range of the vertical axis across the plots; a higher $\bar{n}$ leads to many more false positive counts. The blue data points correspond to a first-order red sideband (RSB1) readout scheme, the orange points to a second-order red sideband (RSB2) and the green points to a third-order red sideband (RSB3) readout. The solid lines are sinusoidal fits to the data. Each data point is the result of 15000 RCS sequences. We simulate 50 independent fringes and show the standard deviation across all 50 realizations as the error bars on each point. We assume we can always couple enough motional quanta (equal to or greater than the RSB order times the trap spacing $\omega_{\textrm{sec}}/2\pi$) to the Be$^{+}$ for a successful readout at a given sideband order, in case of successful He$^{+}$ excitation. Quantum projection noise and false positive counts from Be$^{+}$ (as depicted in Fig. \ref{fig:RSB_order}) are included in these simulations.}
\end{figure}

Fig. \ref{fig:RSB_order_fringes} shows simulated RCS fringes using 1$^{\textrm{st}}$, 2$^{\textrm{nd}}$ and 3$^{\textrm{rd}}$ order RSB readout for a logic Be$^{+}$ ion with $\Bar{n} =$ 0, 0.013 and 0.075. RAP has the potential for near-perfect, coupling-strength-insensitive transfer efficiencies \cite{Gebert_2016, Gebert_2018_corrigendum}. Additionally, we estimate that by carefully tuning the ion trap potentials, the average motional number can be as high as $\bar{n} \approx 20$ on the coupled radial IP mode after a successful He$^{+}$ excitation. We therefore assume in our simulations that we can always couple enough motional quanta to the shared He$^{+}$-Be$^{+}$ radial motional modes for a successful readout of He$^{+}$ excitation using a given RSB order RAP on the Be$^{+}$ ion. The simulation includes quantum projection noise as well as false positive counts due to the logic ion's initial (before excitation and He$^{+}$ radial mode coupling) thermal energy. The plots display a single RCS fringe for each $\bar{n}$ and sideband order, randomly chosen from a set of 50 such fringes. Each data point corresponds to the accumulated counts over 15000 RCS sequences. The error bar on each data point corresponds to the standard deviation on the accumulated counts across all 50 fringes in the set. The solid lines are fits to the data using a sinusoidal model. At $n =0$, the difference in RSB order readout has no effect. As the $\bar{n}$ increases, the advantage of higher-order sideband readout becomes visible as a lower background and better signal contrast.

\begin{figure}[h!]
\includegraphics[width=0.95 \columnwidth]{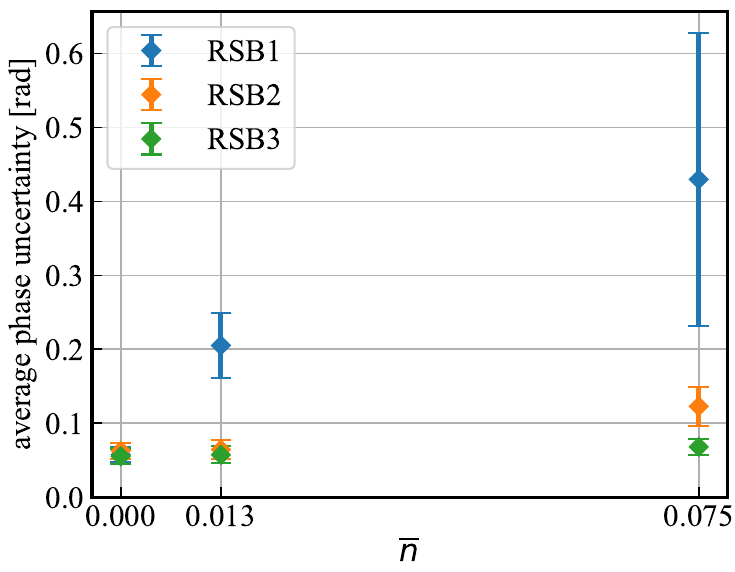}\caption{\label{fig:fit_phase_uncertainty_rsb} Average phase uncertainty (standard deviation) at the transition wavelength on the fitted phase of the data in Fig. \ref{fig:RSB_order_fringes} for all 50 fringes in the set, for different average motional numbers $\bar{n}$, for RSB order readouts 1 (RSB1), 2 (RSB2) and 3 (RSB3) in blue, orange, and green, respectively. Error bars correspond to the standard deviation of the phase uncertainty distributions at each $\bar{n}$ and RSB order for a set of 50 RCS fringes. Phase uncertainty translates into frequency uncertainty in RCS, as explained in detail in section \ref{feasibility section}. At $n=$0, the phase uncertainty for the three RSB orders overlaps. At $\bar{n}=$ 0.075, the phase uncertainty on RSB3 is 84\% smaller than RSB1 and 45\% smaller than RSB2.}
\end{figure}

The transition frequency determination in Ramsey-comb spectroscopy is based on the phase of the sinusoidal fits to the RCS fringes at multiple macrodelays, as is explained in section \ref{feasibility section}. The uncertainty on the phase of the fit is therefore an indicator of the achievable frequency uncertainty. Fig. \ref{fig:fit_phase_uncertainty_rsb} shows the uncertainty on the phase of the fits to the data shown in Fig. \ref{fig:RSB_order_fringes}. The error bars correspond to the standard deviation of the distribution of phase uncertainty for each $\bar{n}$ and RSB combination over the set of 50 fringes. The three points overlap at $n = 0$. At $\bar{n} = 0.013$, the uncertainty on both RSB2 and RSB3 is roughly $\sim 60$ mrad, whereas that of RSB1 is about 3.3 times higher. At $\bar{n} = 0.075$, the uncertainty on the phase of fringes readout with RSB1 is about $430$ mrad, with a large spread, whereas for RSB2 and RSB3 it is around 123 mrad and 68 mrad, respectively, with a significantly lower standard deviation. It is clear from this figure that higher sideband order readout, with its higher signal-to-noise ratio, offers the potential for a significantly lower uncertainty on the transition frequency determined with RCS.

\section{Laser phase noise}\label{laser phase noise}
The simulated RCS fringes presented thus far were produced in the absence of laser phase noise. One of the main challenges in measuring the $1S-2S$ transition in He$^{+}$, however, is the need for phase-coherent radiation in the extreme ultraviolet spectral range. Corsi \textit{et al.} \cite{Corsi} used a Michelson interferometer to measure the pulse-to-pulse phase jitter of an ultra stable metrology frequency comb laser in the near infrared (NIR) at the tens, to hundreds of nanoseconds time scale relevant for an optical Ramsey-type measurement. Phase noise at such short time scales is mainly due to quantum noise in the form of spontaneous emission and intensity fluctuations of the pump laser~\cite{Ansquer_2021, Brandt} that translate into phase noise. Cavity feedback mechanisms typically only become effective on timescales of a microsecond or longer, resulting in an essentially free-running FC for shorter timescales. 

In order to estimate the expected coherence time for such a FC after upconversion to the extreme ultraviolet, Corsi \textit{et al.} up-convert their measured relative phase jitter to the 11th harmonic of their 1550 nm FC (140.9 nm) according to the rules for classical frequency up-conversion given by \begin{equation}\label{harmonic upconversion}
\Delta\phi_{XUV} = q \cdot \delta\xi 
\end{equation} 
with $q = 11$ the harmonic order and $\delta\xi$ is the relative phase jitter of the FC. They find coherence times on the order of 1 microsecond. In contrast, Benko \textit{et al.} reported on the production of an XUV FC with a coherence time exceeding 1 second \cite{Benko}, determined by measuring the XUV FC linewidth using a heterodyne beat between two XUV combs produced by up-converting a common 1070 nm Yb:fiber FC in two independent femtosecond enhancement cavities. The surprising difference between the 1 microsecond coherence time at 140.9 nm reported by Corsi \textit{et al.} and the $>$1 second coherence times at up to 62.9 nm (17th harmonic) reported by Benko \textit{et al.} can be ascribed to two main effects. On the one hand, the femtosecond enhancement cavities used in \cite{Benko} suppress the relative phase jitter to a certain extent, although the cavity finesse is intentionally kept low, thereby limiting its efficacy in this respect. On the other hand, and more significantly, the two XUV combs that are used to determine the coherence time of a given harmonic have correlated phase noise due to the fact that they are derived from the same 1070 nm FC. It is likely that, because of the shared origin, this leads to a reduced phase noise on the beat note. A measurement of the XUV coherence with respect to an independent reference (like a He$^+$ ion) may reveal the presence of considerably more phase noise, closer to that reported (at a significantly longer wavelength) by Corsi \textit{et al.} However, the measurement by Benko \textit{et al.} does provide clear evidence that the phase noise added by the HHG process itself can be very low. Because current technology does not yet enable feedback at very short time scales ($<$1$\mu$s) on FC lasers to suppress phase noise as required for our He$^+$ spectroscopy, an alternative approach is needed. In section \ref{phase noise model section} we propose the use of a high finesse filtering cavity to sufficiently reduce phase noise at the fundamental wavelength prior to upconversion to the XUV. 

\begin{figure}
\centering
\includegraphics[width = \columnwidth]{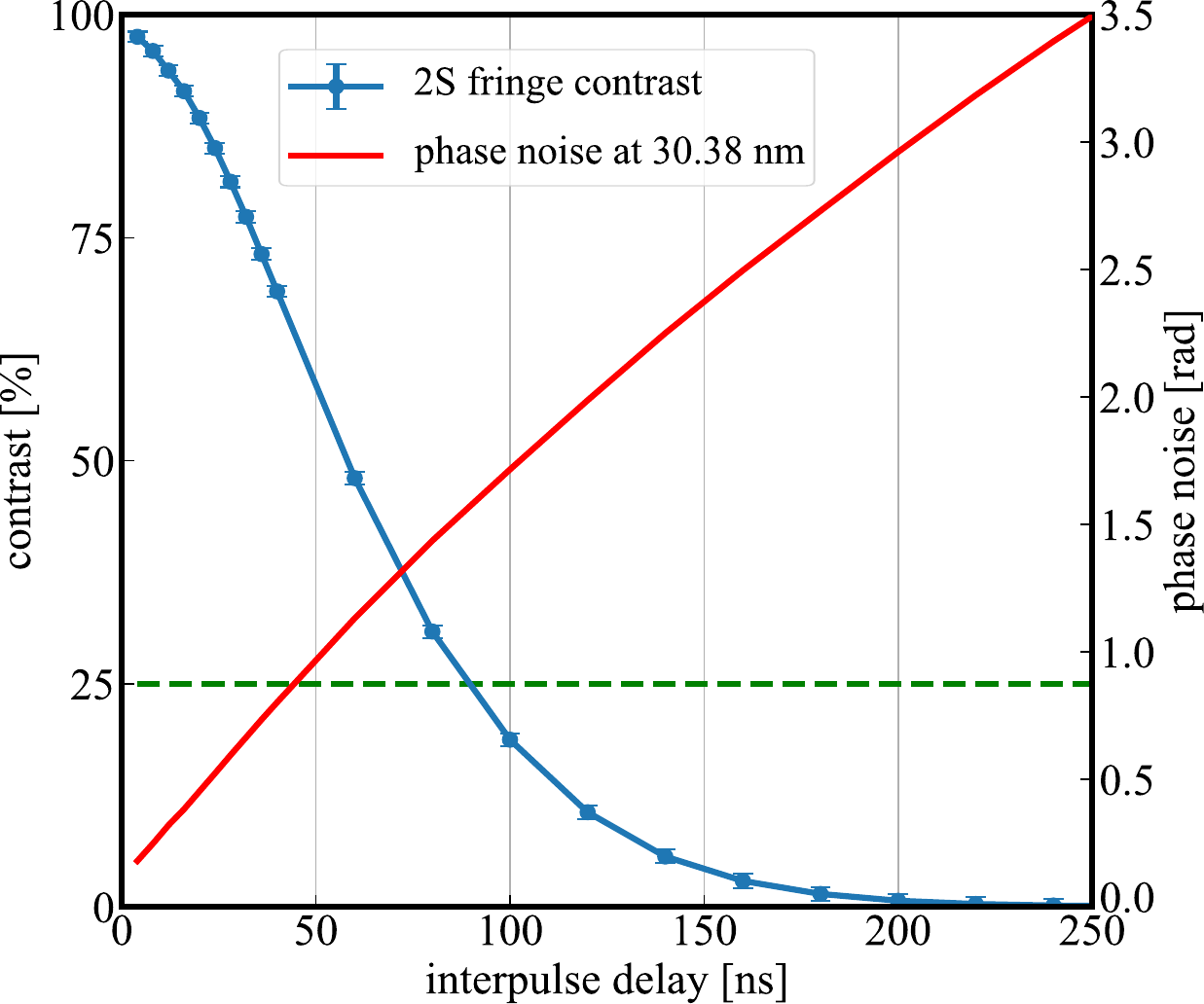}
\caption{{\label{fig:PN_contrast_delay} RCS fringe contrast after incorporating upconverted relative phase jitter data from Ref.~\cite{Corsi} as timing jitter into the TDSE simulation of RCS presented in Fig. \ref{fig:fringes_fig}} (based on 2S population). The red solid line presents the magnitude of this relative phase jitter (right vertical axis) at 30.38 nm incorporated into the RCS signal. The resulting fringes are fitted with a sinusoidal model from which the contrast is extracted, shown as the blue points (left vertical axis, and with small error bars) and connected with a blue solid line. The green dashed line shows the 25\% contrast threshold above which RCS still functions in practice. Without measures to reduce intrinsic FC phase noise, RCS would be limited to macrodelays below 100 ns.}
\end{figure}

To make an initial estimate of the impact of phase noise on Ramsey-fringes in the XUV, we incorporate the most constraining relative phase jitter data presented in Fig. 3 of Ref.~\cite{Corsi} into the TDSE simulations of RCS on a free (untrapped and motionless) He$^{+}$ ion discussed in section \ref{electronic_excitation_dynamics}. The measured phase jitter values are first upconverted from 1550 nm to 30.38 nm according to eq. (\ref{harmonic upconversion}). Our FC has a central wavelength of 1580 nm so we first apply a factor 0.98 to convert the relative phase jitter from Corsi \textit{et al.} to our central wavelength. We then work with a harmonic factor of $q=2\times (25+1)$, where the factor of 2 is due to the upconversion from 1580 nm to 790 nm, and the factor of $25+1$ is due to the fact that we drive the $1S-2S$ 2-photon transition in He$^{+}$ with one 790 nm photon and one photon from its 25$^{\textrm{th}}$ harmonic at 32 nm, effectively the 26$^{\textrm{th}}$ harmonic. The upconverted relative phase jitter is translated into timing jitter and incorporated into the RCS fringes of Fig. \ref{fig:fringes_fig}. The contrast of the resulting fringes based on He$^{+}$(2S) population and for He$^{2+}$ produced from the 2S level by the second RCS pulse is shown in Fig.~\ref{fig:PN_contrast_delay}. Without the use of noise suppression techniques, and in the absence of any other sources of noise, false positive counts or motional level dephasing, RCS would be limited to interpulse delays below 100 ns or half an ion secular period ($25\%$ contrast threshold), thereby severely limiting the frequency resolution of the technique. This estimate motivates the development of a phase noise model and of phase noise filtering that can be incorporated into simulations of the excitation dynamics of trapped He$^{+}$.

\subsection{Phase noise model}\label{phase noise model section}
In order to study the impact of phase noise in the XUV on RCS of trapped ions, we perform a Monte Carlo simulation of the relative phase between two FC pulses that can be incorporated into the total simulation of the experiment. We propose to reduce the FC noise with a vacuum high-finesse resonator that effectively produces a running average of the FC pulses to reduce the phase noise on the relevant time scale of up to a few microseconds. In order to simulate the effect of such a filtering cavity, we first develop a heuristic model that mimics the pulse-to-pulse phase jitter measured by Corsi \textit{et al.} \cite{Corsi} on a FC (Menlo Systems ULN) similar to the one we will use for RCS on He$^{+}$. 

\begin{figure}
\includegraphics[width=\columnwidth]{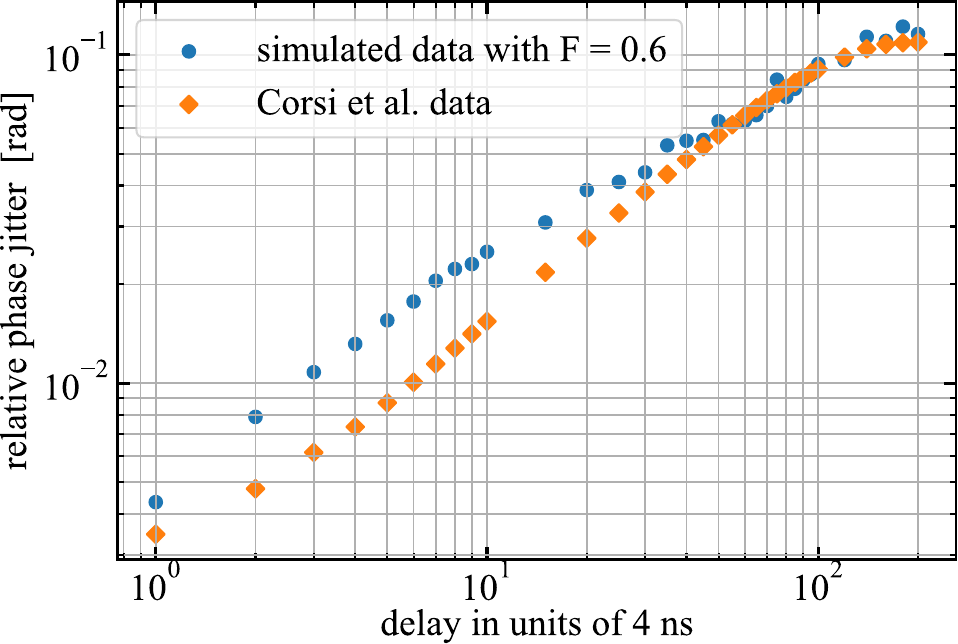}
\caption{\label{fig:f6_phase_noise} Simulated relative phase jitter (blue circles) with a correlation factor of $F=0.6$ and data from Fig. 3 of Ref.~\cite{Corsi} for the maximum FC amplification setting of 380 mW (orange diamonds). Our model overestimates the phase noise at timescales below 200 ns but agrees well with the measured data between 200 and 800 ns. As shown in Fig. \ref{fig:macrodelay_fig}, the minimum interpulse delay to avoid destructive interference between ion motional levels in RCS on trapped ions is 200 ns.}
\end{figure}

We prepare a set of correlated phase noise values $\delta \xi = \{\xi_{n}\}$, whose elements are generated according to
\begin{equation}\label{phase noise model}
    \xi_{n} = \textrm{random}\left[\delta \eta(\sigma)\right] + F\times \xi_{n-1}
\end{equation}
where $\xi_{0}=0$ and $F$ is a correlation strength factor. In the equation, $\textrm{random}\left[\delta \eta(\sigma)\right]$ is a randomly sampled point from a normal distribution $\delta \eta(\sigma)$ with a standard deviation of $\sigma= 0.98 \times 2 \times 3.5$ mrad. We find that $F = 0.6$ approximately reproduces the phase noise as a function of delay of Ref.~\cite{Corsi} as shown in Fig. \ref{fig:f6_phase_noise}.


In order to simulate pulses out-coupled from the filtering cavity, we first define the electric field of pulse $j$ in the cavity as a cosine carrier modulated with a Gaussian envelope 
\begin{equation}
     E_{j}(t) =  R_{A}^{2j}e^{-at^{2}}\cos(\omega_{c}t+ \Phi_{j})
\end{equation}
where $R_{A}$ is the electric field amplitude reflectivity of a cavity mirror, allowing us to account for amplitude decay over $j$ cavity round trips. The exponential is a Gaussian envelope with $a = 2\textrm{ln}(2)/\tau_{\textrm{FWHM}}^{2}$ for a full-width at half-maximum pulse duration of $\tau_{\textrm{FWHM}}$. To generate pulse $p_{i}$, with $i$ indexing the RCS pulse number (1 or 2) out-coupled from the cavity, we add up the electric field of each individual pulse in the cavity according to
\begin{equation}\label{cavity equation}
    p_{i} = \sum_{j} E_{j}(t)
\end{equation}
where 
\begin{multline}\label{cavity_pulse_number}
   j = 
\begin{cases}
    \Delta N \textrm{ to } \Delta N + \gamma & \textrm{for } i = 1 \\
 \textrm{0 to } \gamma & \textrm{for } i = 2
\end{cases}
\end{multline}
where $\Delta N$ is the macrodelay in units of the laser repetition period (8 ns), and $\gamma$ is the number of pulses which interfere in the cavity, in steady state operation. We determine $\gamma$ by finding the number of cavity pulses beyond which the addition of more pulses has no signficant effect on the resulting pulse $p_{i}$. For a finesse of 3000 we find $\gamma = 5000$ to be sufficient. The pulses are simulated over a time span of 2 picoseconds around the peak of the pulses, with a temporal resolution of 10 attoseconds. The phase term $\Phi_{j}$ is given by
\begin{equation}
    \Phi_{j} = \xi_{j} + jM + \kappa_{j}
\end{equation}
where $\xi_{j}$ is given by equation (\ref{phase noise model}). $M$ is a factor accounting for mismatch between the repetition period of the comb $\tau_{\textrm{rep}}$ and the cavity round trip time $\tau_{\textrm{cavity}}$ given by 
$M = \omega_{c}(\tau_{\textrm{rep}}-\tau_{\textrm{cavity}})$. Because we assume tight locking of the cavity this term is taken to be zero. $\kappa_{j}$ is the phase acquired in the cavity due to second-order dispersion
\begin{equation}
    \kappa_{j} = \frac{ 2k^{\prime\prime} j t^{2}}{\frac{1}{b^2}+(2k^{\prime\prime}j)^{2}}
\end{equation}
where $k^{\prime\prime}$ is the group delay dispersion (GDD) per reflection of the mirrors, $j$ the number of round trips in the cavity, and $b = 2\textrm{ln}(2)/\tau_{\textrm{FWHM}}$.

We prepare a complex field corresponding to eq. (\ref{cavity equation}) by filtering out its negative frequency components. The phase of the field can thus be determined easily at every point in time. We sum over the phases at each time point to define an average phase per pulse and thereby a relative phase for pulses pairs out-coupled from the cavity and spaced $\Delta N$ apart. For our simulations we take an initial pulse duration $\tau_{\textrm{FWHM}} = 150$ fs and a GDD of $2 \times 7$ fs$^{2}$ per cavity roundtrip, corresponding to readily available commercial mirrors suitable for our bandwidth of approximately 10 nm centered at 790 nm. 

\begin{figure}
\centering
\includegraphics[width = \columnwidth]{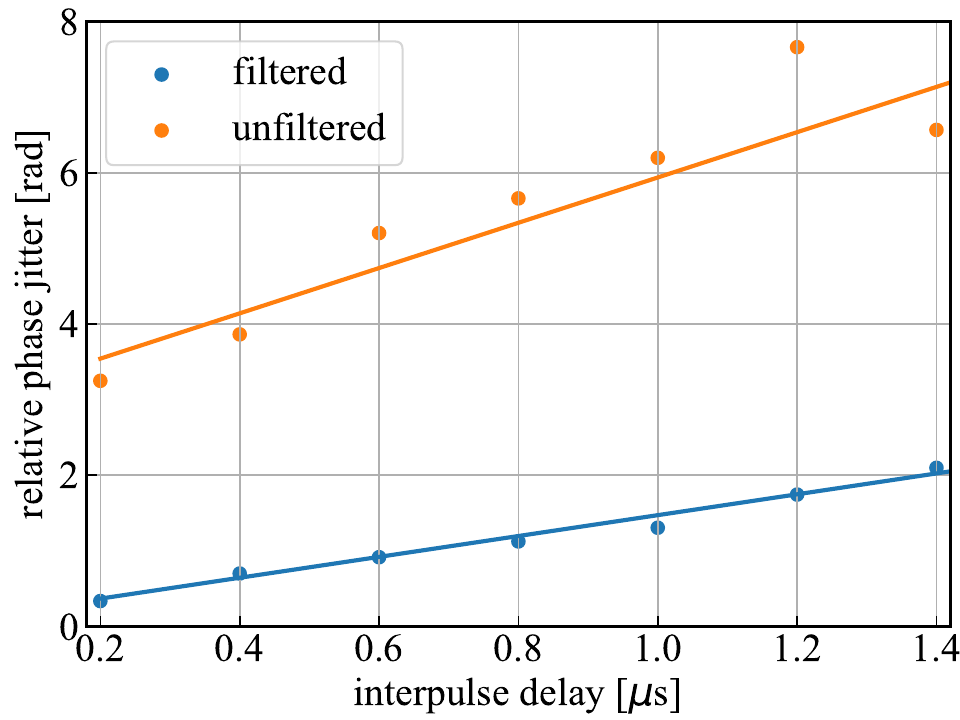}
\caption{\label{fig:PN_filtered_unfiltered} Simulated phase jitter between two pulses upconverted to the transition wavelength of 30.38 nm based on cavity-filtered and unfiltered fundamental pulses from the FC, calculated at multiples of the 200 ns ion secular period, in the absence of technical noise (e.g. amplification noise). The simulation is performed with a cavity finesse of 3000. The blue (orange) circles are the simulated data for filtered (unfiltered) pulses and the solid lines are linear fits to the data.}
\end{figure}

Fig. \ref{fig:PN_filtered_unfiltered} shows the standard deviation of the relative phase jitter distribution of two 790 nm pulses at multiples of 200 ns with and without a filtering cavity, produced by our simulation for a cavity with a finesse of 3000, corresponding to a mirror reflectivity of 0.99895. In both cases the relative phase jitter grows linearly but the slope of the unfiltered phase jitter is 54\% steeper than that of the filtered phase jitter and the curve has a much larger offset. At 200 ns the cavity noise suppression factor obtained from the fits to the data is 9.6 while at 1.4 $\mu$s it is 3.5. 


\begin{figure}
\includegraphics[width = 0.75 \columnwidth]{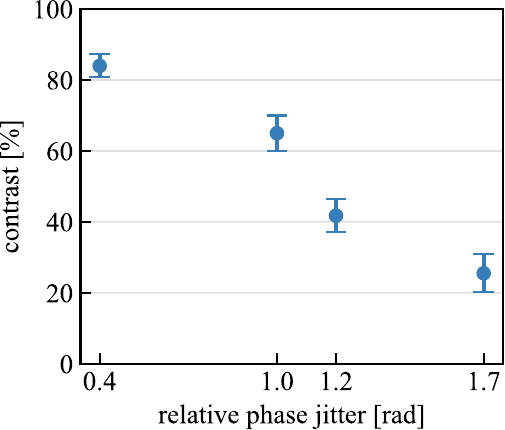}
\caption{\label{fig:contrast_multiples25} RCS fringe contrast as a function of phase noise (see Fig.~\ref{fig:PN_filtered_unfiltered}) for the $1S-2S$ transition in He$^{+}$. Each point is an average over 50 independent RCS simulations, for a range of phase noise values at the transition wavelength of 30.38 nm. For this graph we assume an ion in an $n=0$ Fock state with no background counts, quantum projection noise, readout with QL for an ion at $\bar{n}$=0 (so no false positives), and perfect synchronization between the ion secular period and the interpulse delay. The error bars correspond to the standard deviation of the contrast distributions per data point.}
\end{figure}


Fig. \ref{fig:contrast_multiples25} shows the average contrast of 50 simulated RCS fringes as a function of relative phase jitter at 30.38 nm for an ion in an $n=0$ Fock state,  with quantum projection noise, readout with QL, and with perfect $\tau_{\textrm{rep}}$ mod$\Delta N = 0$ synchronization. The error bars show the standard deviation of the contrast distributions. Because a perfectly cold ion does not suffer from false positive counts, this loss of contrast is due to a decrease in amplitude modulation of the fringes with increasing phase noise. A lower contrast therefore indicates a higher uncertainty on the sinusoidal fit to the fringes. A simulation with the same parameters but no filtering cavity shows that phase noise in the XUV leads to zero contrast even in the case of synchronization. 

Because the achievable frequency uncertainty in Ramsey-type measurements is a function of the maximum interpulse delay, it is clear that XUV phase noise represents the most limiting factor to higher spectral precision. Based on our findings, we conclude that with a moderate finesse of 3000 for the filter cavity, Ramsey pulse delays up to 1 $\mu$s are possible. For a cavity finesse of 45000, we expect delays of up to 6 $\mu$s to be possible with sufficient fringe contrast, as described in section \ref{feasibility section}.

\section{Experimental feasibility of the $1S-2S$ transition in He$^{+}$}\label{feasibility section}

\begin{figure*}
\includegraphics[width= 0.72\textwidth]{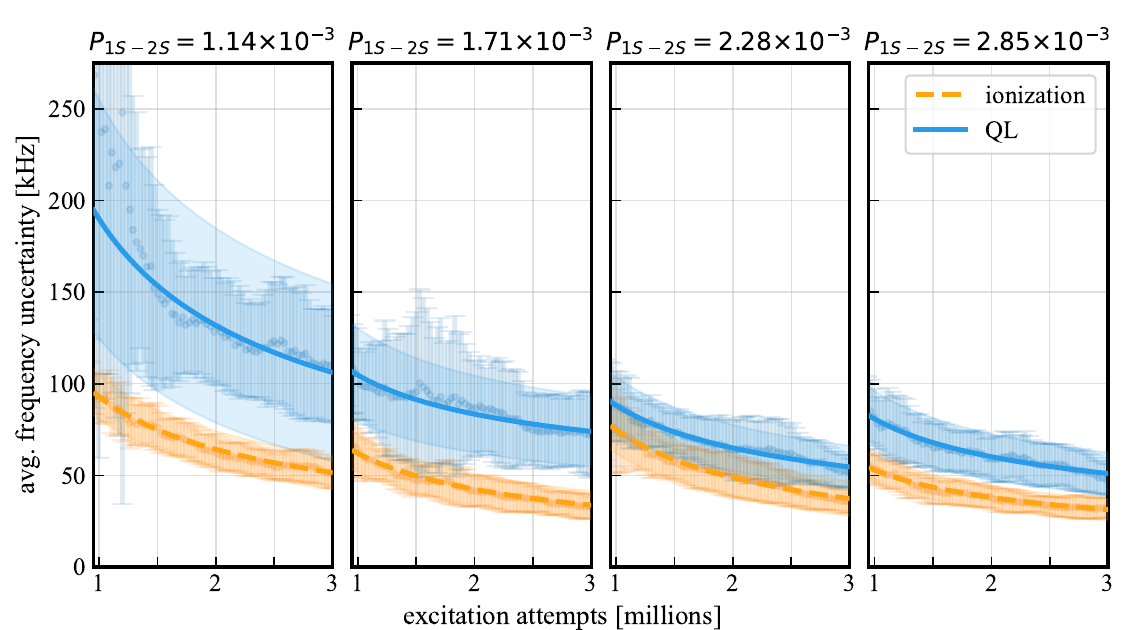}\caption{\label{fig:delta_f_shots} Simulated frequency uncertainty as a function of excitation attempts for single-pulse electronic excitation probabilities of 1.14$\times 10^{-3}$ to 2.85$\times 10^{-3}$, corresponding to a 2-, 3-, 4-, and 5-fold increase in XUV production relative to the XUV intensity estimated in Ref.~\cite{Grundeman}. The lines show fits to the simulated data (shown in darker colors in the background), and the solid bands show the $\pm \sigma$ uncertainty interval. Blue solid lines and bands correspond to QL readout while dashed orange lines and orange bands correspond to ionization readout.}
\end{figure*}

\begin{figure}
    \includegraphics[width=0.87\columnwidth]{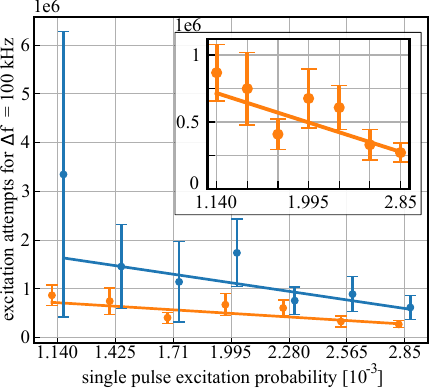}
    \caption{Number of excitation attempts required to reach a target frequency uncertainty of 100 kHz as a function of the single pulse $1S-2S$ excitation probability, extracted from the fits to the data presented in Fig. \ref{fig:delta_f_shots}, fitted with a linear model as a first-order approximation to its true functional form, and to guide the eye. Data in blue (orange) corresponds to QL (ionization) with RSB2 readout. The inset shows a zoom-in to the ionization data for better visibility. A slight horizontal shift is introduced between the two data sets for improved visibility.}
    \label{fig:shots_prob}
\end{figure}

We now combine the motional excitation dynamics and XUV phase jitter simulations to investigate the underlying trends in the parameter space associated with the use of a filtering cavity with a finesse of 3000. We take an initial thermal logic and spectroscopy ion motional states distribution of $\Bar{n} = 0.075$ as a realistic yet experimentally ambitious goal. This combination of cavity finesse and $\bar{n}$ leads to a minimum achievable frequency uncertainty of 100 kHz. We then provide feasibility estimates for a target uncertainty of 10 kHz for an RCS measurement of the $1S-2S$ transition in He$^{+}$. The transition frequency determination in RCS is performed in the time domain with a fitting algorithm based on the phase of the individual RCS fringes at different macrodelays (see Ref.~\cite{Jonas_theory}). The minimum required measurement is based on two fringe pairs at different macrodelays, each producing a phase difference. We can estimate the achievable frequency uncertainty based on a quadratic addition of the uncertainty (standard deviation) of the two fitted phase differences $\sigma_{\phi_{N_{i}}}$, and the time difference $\tau$ between the two selected macrodelays $N_{i=1}$ and $N_{i=2}$ according to
\begin{equation}\label{freq_uncertainty_eq}
    \Delta f = \frac{1}{2\pi\tau}\sqrt{\sigma_{\phi_{N_{1}}}^{2}+\sigma_{\phi_{N_{2}}}^{2}}
\end{equation}
where $\tau = \frac{1}{f_{\textrm{rep}}}(N_{2} - N_{1})$ with $N_{2}>N_{1}$. $\sigma_{\phi_{N_{i}}}$ is the standard deviation on the estimated phase from the non-linear least squares fit of the RCS fringe.

The relative phase jitter used for these simulations corresponds to a FC laser filtered in a linear filtering cavity with a finesse of 3000, whose free spectral range is perfectly matched to the frequency comb teeth spacing. In addition to experimental noise, the simulations used for this feasibility estimate also include the effect of quantum projection noise. 

The RCS laser system is based on selective amplification of FC pulses in an optical parametric amplifier (OPA) (see section \ref{RCS_technique}). Intensity fluctuations in the pump beam of the OPA can lead to phase fluctuations in the amplified FC pulses. For that reason, in addition to laser phase noise discussed in \ref{phase noise model section}, we consider phase noise due to the optical parametric amplification process at the level of 15 mrad at 790 nm, as expected from simulations of our newly-built lithium triborate (LBO) based OPA. These noise sources are combined and upconverted to the transition wavelength. Noise in the high harmonic generation process and ac-Stark shift noise due to intensity fluctuations constitute negligible contributions to the total pulse-to-pulse phase jitter, and are therefore not considered in this study  \cite{Grundeman}. 

We carry out simulations for both the QL readout scheme described in section \ref{QLS_sec}, and the state-selective ionization scheme in which an ionization pulse ionizes He$^{+}(2S)$ to He$^{2+}$. In that latter case, the trapping potential can be adjusted such that doubly-ionized helium is not trapped so that it will leave the trap. The loss of He$^{2+}$ from the ion trap induces a change in the Be$^{+}$ position which can be detected by monitoring the fluorescence of Be$^{+}$. For ionization readout, we consider an ionization pulse which comes a few nanoseconds after the RCS excitation sequence so that it can be optimized independently from the excitation pulses for a maximum contrast between signal and background ions.

For the QL scheme, we carried out simulations using both first- and second-order RSB readout. RAP population transfer would also allow for the use of higher RSB orders although transfer efficiencies become increasingly motional state-dependent as the RSB order increases. Readout using a first-order RSB for an ion with $\Bar{n} = 0.075$ proved to have a signal-to-noise ratio that is insufficient for a precision determination of the transition frequency. Therefore, only second-order RSB RAP readout is considered from here on. We ran simulations using $N_{1} = 25$ (200 ns), and $N_{2} = $ 100 (800 ns) and 125 (1 microsecond). For the chosen FC laser phase noise conditions, we find that the decrease in uncertainty using $N_2$ = 125 instead of $N_2 =$ 100 is almost exactly countered by the detrimental effect of the increase in laser phase noise with delay, leading to a net zero improvement in frequency uncertainty. We therefore only present results for $\tau = 600$ ns, corresponding to $N_1 =$ 25 and $N_2$ = 100, as that is experimentally easier to achieve.


From our prior experimental and numerical study of the excitation dynamics of He$^{+}$ excited with a single two-color (NIR + XUV) pulse \cite{Grundeman}, we determine that below a NIR intensity of about $0.2 \times 10^{14}$ W/cm$^2$, multiphoton ionization of He$^{+}(2S)$ to He$^{2+}$ with the NIR field becomes negligible (approximately $4.2 \times 10^{-4}$). This NIR intensity translates into a $1S-2S$ electronic excitation probability $P_{1S-2S}$  of $0.57 \times 10^{-3}$ per pulse. We simulate an up to 5-fold increase in the $1S-2S$ electronic excitation probability, as could be achieved by an increase in the HHG yield of the same magnitude, while keeping the NIR intensity the same. In our prior work \cite{Grundeman}, the generation of high harmonics was not fully optimized. We later observed that relatively simple changes in experimental parameters could produce a significantly higher HHG yield, such as further increasing the backing pressure of the valve injecting the HHG medium (argon) into the vacuum chamber. Additionally, our setup features an adjustable iris after the HHG chamber (see Fig. \ref{fig:experimental_setup}) which allows us to reduce the NIR intensity without cutting into the XUV. We therefore believe an up to a 5-fold increase in HHG yield to be realistic. At low excitation rate, state-selective ionization will still work for good contrast RCS signal by simply measuring for a longer time as ionization can be arranged such that there are (almost) no background counts. For QL this is different because it is much more prone to false-positive counts. In that case we find that the signal-to-noise ratio of data produced with $P_{1S-2S} < 1.14 \times 10^{-3}$ is not high enough to reach a frequency uncertainty of 100 kHz. We therefore only consider data produced with $P_{1S-2S} \geq 1.14 \times 10^{-3}$, corresponding to a 2-fold or higher HHG yield increase compared to our previous study~\cite{Grundeman}. Note that in~\cite{Grundeman} the actual excitation probability was closer to $10^{-4}$ due to the non-optimal experimental conditions based on an atomic helium beam.

We estimate the uncertainty on the transition frequency according to eq. (\ref{freq_uncertainty_eq}) for a set of 10 independent RCS fringe pairs as a function of the number of excitation attempts per fringe pair (RCS experiment runs). We average the set of 10 frequency uncertainty values per excitation attempts, find their standard deviation and use these data as input for a weighted least squares fit using the model
\begin{equation}
\Delta f(n) = \frac{a}{\sqrt{n}} + b    
\end{equation}
where $n$ is the total number of excitation attempts for a given average frequency uncertainty, and $a$ and $b$ are free parameters. Fig. \ref{fig:delta_f_shots} shows the data in the background, with the resulting fits and $\pm 1\sigma$ uncertainty band in the foreground. The error bars on the data points correspond to their standard deviation. The solid blue line corresponds to QL readout, and the dashed orange line to ionization readout. The curves are shown for $1S-2S$ electronic excitation probabilities $P_{1S-2S}$ of 1.14, 1.71, 2.28, and 2.85 $\times 10^{-3}$ which correspond to a 2-, 3-, 4-, and 5-fold increase in the XUV yield, respectively. Curves corresponding to $P_{1S-2S}$ = 1.425, 1.995 and 2.565 $\times 10^{-3}$ are also considered, although not shown in Fig. \ref{fig:delta_f_shots}.

\begin{figure}
\includegraphics[width= \columnwidth]{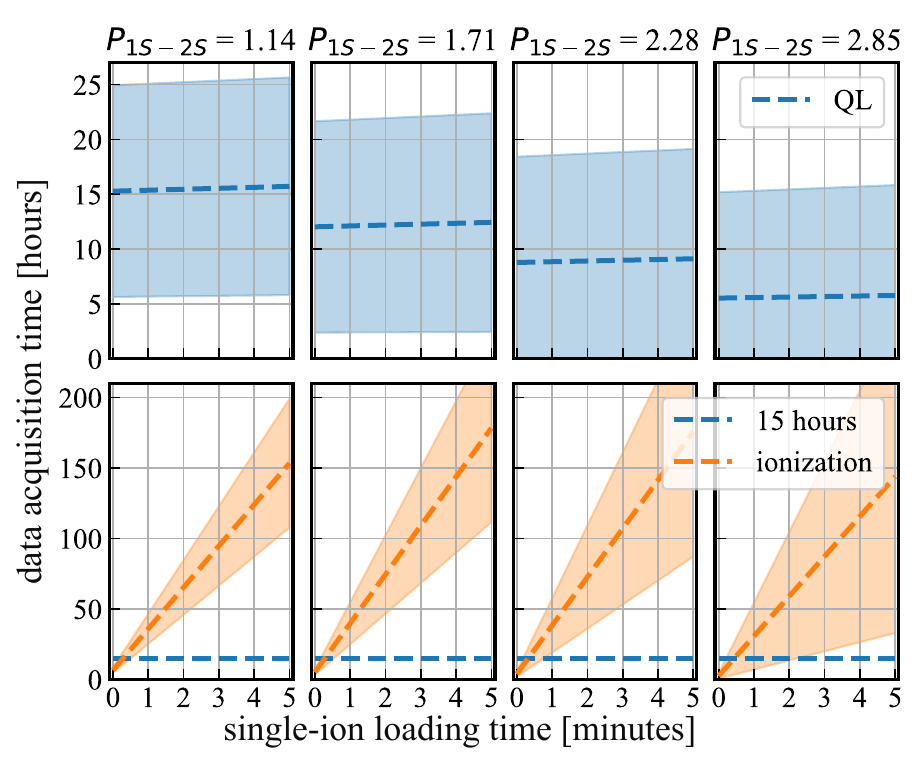}\caption{\label{fig:data_acq_time} Total data acquisition time needed to reach 100 kHz frequency uncertainty as a function of the single-ion loading time for different single-pulse $1S-2S$ electronic excitation probabilities $P_{1S-2S}$ (given in units of $10^{-3}$). Upper row in blue QL readout, lower row in orange ionization readout. The solid bands represent $\pm 1\sigma$ uncertainty (determined from the variance in the simulation results). The acquisition time is calculated using eq. (\ref{data_acq_time_ion}) and eq. (\ref{data_acq_time_QL}), with the number of shots to reach 100 kHz, $S$, in $N_{2S}$ and $N_{0}$ determined from the linear fit to the data in Fig.\ref{fig:shots_prob}. The dashed blue line in the lower row of plots shows the 15 hour mark for comparison, corresponding to approximately the longest data acquisition time under a QL readout.}
\end{figure}

As can be seen from the QL curves, as $P_{1S-2S}$ increases, the required number of excitation attempts (and the associated uncertainty) first decreases significantly due to an increase in the SNR and then enters an asymptotic regime. The same effect is present in the ionization readout curves, albeit much less dramatically so, as the ionization SNR only depends on experimental noise and is insensitive to false positive counts. The range of excitation attempts is bounded on the lower end by the necessary SNR to fit the RCS data with a sinusoidal model at all of the $P_{1S-2S}$ values here considered. On the upper end, the range is bounded by the difficulty in doing measurements with more than 3 million excitation attempts, and by computational cost in the case of simulated data. We observe that the lowest achievable frequency uncertainty is lower using ionization readout for all values of $P_{1S-2S}$. However, the difference becomes significantly smaller as the excitation probability increases. Although the necessary excitation attempts for a given target uncertainty may be lower for ionization readout, the total measurement time is higher due to the required reloading of an ion after every successful excitation to the 2S state, as is discussed below and shown in Fig. \ref{fig:data_acq_time}. 

From the least-squares estimate of the fit parameters $a$ and $b$, we determine the number of excitation attempts necessary for a frequency uncertainty of 100 kHz for a range of $P_{1S-2S}$ values with the results shown in Fig. \ref{fig:shots_prob}. The error bars correspond to the standard deviation on the required excitation attempts, calculated using the standard deviation of the fit parameters. The inset is a zoom-in to the ionization data. The data is fitted with a linear model as a first-order approximation to the true functional form. We find that the number of required excitation attempts goes down roughly linearly with $P_{1S-2S}$, and that the error on the fit also goes down with $P_{1S-2S}$ for both readout types. As in Fig. \ref{fig:delta_f_shots}, the number of required shots for the same frequency uncertainty is consistently lower for ionization readout. 

Using the estimated number of excitation attempts to reach 100 kHz uncertainty, $S$, shown in Fig. \ref{fig:shots_prob}, we can estimate the total data acquisition time corresponding to each readout type. We assume that the 30 ms between each double-pulse laser excitation attempt (from the overall 30 Hz repetition rate of the RCS laser system) is sufficient to restore the initial conditions for the next attempt. We consider the data acquisition time per attempt in the absence of 2S excitation $t_{S} =  1 \textrm{ min.}/1800 \textrm{ attempts}$, the ion reloading time $t_{ion}$, the He$^{+}(2S)$ multi-photon ionization probability of $\Gamma_{2S_{ion}} = 4.2 \times 10^{-4}$ from the RCS pulses themselves (only important for the QL scheme), and the fractional excitation probability $\alpha = $ 2S atoms/(total excitation attempts). We determine the number of successful excitation attempts $N_{2S} = \alpha \times S$ and the number of unsuccessful excitation attempts $N_{0} = S \times(1-\alpha)$.  Alltogether, it leads to a total data acquisition time of 


\begin{equation}\label{data_acq_time_ion}
    t_{data, i} = N_{2S} \times t_{ion} + N_0 \times t_S
\end{equation}
for ionization readout and for QL readout, one finds
\begin{multline}
\label{data_acq_time_QL}
    t_{data, QL} = N_{2S} \times \Gamma_{2S_{ion}} \times t_{ion} \\
    + (N_0 + N_{2S}(1-\Gamma_{2S_{ion}})) \times t_S
\end{multline}
where we assume vacuum conditions in the experimental chamber to be good enough to neglect background collisions with the trapped ions that require reloading. We determine the uncertainty on the estimate of the data acquisition time by considering the standard deviation of the estimated number of shots. The results are shown in Fig. \ref{fig:data_acq_time}, where we find that despite requiring more shots for the same target uncertainty, QL readout can lead to significantly shorter data acquisition times (e.g. over an order of magnitude), and increasingly so for higher electronic excitation probabilities and longer ion reloading times. This is reflected in the crossing points for the ionization and QL curves which are at 17, 11, 8, and 6 seconds for $P_{1S-2S}$ 1.14, 1.71, 2.28 and 2.85 $\times 10^{-3}$, and respectively. From this estimate, we find that a measurement at the 100 kHz uncertainty level could be reached within 25 hours of data acquisition for a QL readout by a two-fold increase in the HHG yield relative to the experiment presented in \cite{Grundeman}. By rewriting eq. (\ref{data_acq_time_ion}) as $t_{data,i} = S\alpha(t_{i}-t_{S}) + St_{S}$, the relatively small effect of increasing $P_{1S-2S}$ on $t_{data,i}$ for ionization readout can be ascribed to the simultaneous decrease in $S$ and increase in $\alpha$. In other words, for ionization readout, which does not suffer from false positive counts, reaching a given target uncertainty is simply a matter of getting enough 2S counts over an RCS fringe. If the ion reloading time is much greater than the time to get a 2S count, then an increase in the excitation probability, and thereby a decrease in the time to get a 2S count, has a relatively minor impact on the data acquisition time. For QL readout, an increase in the electronic excitation probability has a significant impact as it directly leads to a significantly higher signal-to-noise ratio.

\subsection{Reaching 10 kHz frequency uncertainty}

A precision spectroscopic measurement of the 1S-2S transition of He$^{+}$ with 100 kHz uncertainty ($\sim 10^{-11}$ relative uncertainty) would represent a great step in the advancement of XUV spectroscopy. However, in order to use a measurement of the $1S-2S$ transition frequency in He$^{+}$ for tests of quantum electrodynamics or for a competitive determination of fundamental constants, an uncertainty at the tens of kHz level is desirable. It is clear that one of the central challenges to carrying out a measurement with higher precision is the large laser phase noise at the transition wavelength. 

The relative phase jitter of the FC pulses could be improved by working with a higher-finesse cavity than the 3000 considered in these simulations, allowing for significantly longer maximum interpulse delays and thus a longer delay $\tau$ between the two fringes used for a frequency determination. One of the main challenges in using high finesse cavities for ultrashort pulses lies in the difficulty of precisely matching the frequency comb teeth in the frequency domain to the narrow eigenmodes of the cavity. In the time domain, this translates to matching the repetition frequency of the FC $f_{\textrm{rep}}$ to the free spectral range (FSR) of the cavity. Locking the cavity FSR to $f_{\textrm{rep}}$ also requires careful adjustment of the FC offset frequency $f_{0}$ to the effect of mirror dispersion of the cavity. The stronger higher-order dispersion is from mirror coatings, the harder it is to fulfill the resonance condition over a large bandwidth, thereby reducing the bandwidth of the filtered pulses and increasing the Fourier-limited pulse duration. Despite these technical challenges, high-finesse (28000 to 45000) optical enhancement cavities for infrared and near-infrared pico and femtosecond pulses with free spectral ranges between 76.4 and 216.66 MHz have been reported \cite{Lu, Lu2, Borzsonyi}. 

By using the data sets in Fig. \ref{fig:delta_f_shots} and leaving $\tau$ in eq. (\ref{freq_uncertainty_eq}) as a free parameter, we estimate the $\tau$ value that yields a frequency uncertainty around 10 kHz for 3 million excitation attempts. Furthermore, we assume that the data is acquired with an XUV phase noise of approximately 1.3 radians; the maximum practical upper bound (and corresponding to the phase noise at 800 ns pulse delay for a cavity finesse of 3000 as we simulated earlier). We present the results in Table \ref{QL_ion_table} below.

\begin{table}[h]
\centering
\renewcommand{\arraystretch}{1.2} 
\begin{tabular}{|c || c  c  c  c|} 
\hline
$P_{1S-2S}$ & 1.14$\times 10^{-3}$ & 1.71$\times 10^{-3}$ & 2.28$\times 10^{-3}$ & 2.85$\times 10^{-3}$ \\ [0.5ex]
\hline
ionization & 3 $\mu$s & 2 $\mu$s & 2.1 $\mu$s & 2 $\mu$s\\
\hline
QL & 6 $\mu$s & 4.5 $\mu$s & 3.5 $\mu$s & 3.2 $\mu$s \\ [1ex] 
\hline
\end{tabular}
\caption{Necessary delay $\tau = \frac{1}{f_{\textrm{rep}}}(N_{2} - N_{1})$ to reach 10 kHz uncertainty with 1.3 rad phase noise in the XUV, corresponding to the data set in Fig.~\ref{fig:delta_f_shots}, for different electronic excitation probabilities $P_{1S-2S}$ per RCS pulse sequence.}
\label{QL_ion_table}
\end{table}

\begin{figure}
\includegraphics[width =  0.77\columnwidth]{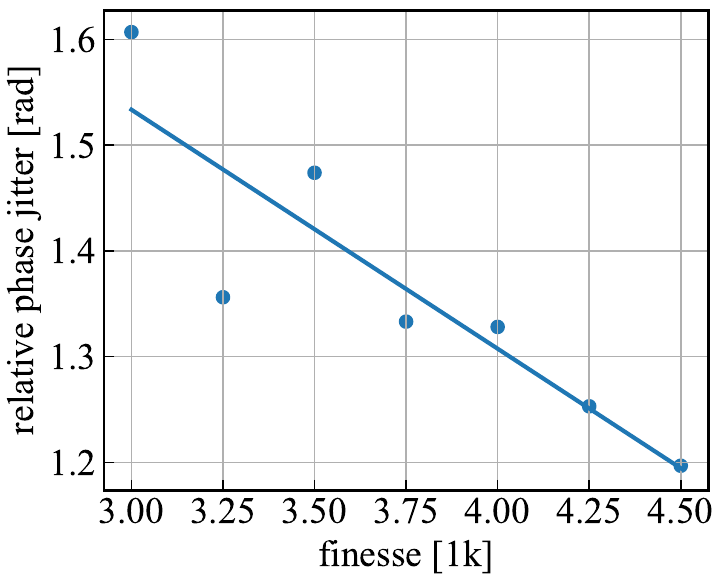}
\caption{\label{fig:PN_fin_1us} Relative phase jitter at 30.38 nm as a function of finesse (in units of 1000) for an interpulse delay of 1 microsecond. Each point is the standard deviation of a set of 500 relative phase samples produced by our simulation.}
\end{figure}

Our laser phase noise simulation does not include the effect of feedback on the laser cavity length, which starts to have an effect only at delays beyond 1 $\mu s$, which is the minimum duration over which the electronics for such a stabilization technique start to work \cite{Corsi}. The relative phase jitter will reach an initial maximum value at 1 $\mu s$, after which it will plateau and increase much more slowly. It is therefore reasonable to expect that in the range of the 2 to 6 $\mu s$ delays that yield 10 kHz uncertainty presented in table \ref{QL_ion_table}, the laser phase noise will be approximately the same as at 1 $\mu s$ delay. In order to determine the cavity finesse $F$ that would allow us to obtain 1.3 radians XUV phase noise at 1 $\mu s$ delay, and thus a frequency uncertainty around 10 kHz for both readout types, we simulate the XUV laser phase noise at 1 $\mu s$ delay for various finesse values and show the results in Fig. \ref{fig:PN_fin_1us}. The linear decrease of the relative phase jitter with finesse is expected due to the linear dependence of the cavity lifetime on finesse which goes as $\tau_{cav} = F/\pi\nu_{FSR}$ where $\nu_{RSR}$ is the cavity free spectral range. We find that a relatively modest increase in cavity finesse from 3000 to approximately 4000 would already enable XUV coherence times $\geq 1 \mu$s, and would allow for a measurement with 10 kHz uncertainty.

\subsection{Single-pulse normalization}

According to the presented simulations and estimates, a measurement campaign would always span at least multiple days to obtain an uncertainty on the $1S-2S$ transition of 100 kHz or better. During this time, essential parameters such as the overlap of the XUV beam with the He$^{+}$ ion or the intensity of the XUV and NIR beam (or the ion temperature) will likely vary. This would lead to a distortion and reduced contrast of the RCS signal. We propose to circumvent the influence of these aspects in two ways. First by scanning the pairs of RCS fringes in a random way so that drifts during the measurement influences all RCS fringes in a similar way, as we have shown before in \cite{Dreissen,Charlaine}. Secondly, by alternating a measurement between one and two excitation pulses, where the one-pulse excitation probability is used to normalize the coherent two-pulse RCS signal. 

The single-pulse signal is a measure for the excitation probability (which includes aspects of the NIR and XUV pulse intensities and the overlap with the ion), while the two-pulse signal measures the coherent evolution (interference). In this manner, signal from many days and different conditions can be combined to reconstruct an RCS signal that is independent of many technical parameters of the experiment. Of course, a varying overlap of the NIR beam would also induce a varying ac-Stark shift. However, the one-pulse signal can also be used to monitor and correct excitation intensity and overlap with the He$^+$ ion to limit such effects. Note that using the single-pulse normalization technique implies that the data acquisition time is twice as long as that shown in Fig. \ref{fig:data_acq_time} in which this additional step was not included. 

\section{Conclusion and outlook}

In this article, we numerically investigated key aspects of extreme-ultraviolet (XUV) spectroscopy on trapped ions outside the Lamb-Dicke regime. This allowed us to better understand the parameter space in which precision XUV spectroscopy on single trapped ions is possible, considering the constraints imposed by such a system, notably XUV laser phase noise and the quantization of motional levels in an ion trap. We developed a phase noise model that allows us to study the relative phase jitter of frequency comb pulses at time scales relevant for optical Ramsey-type measurements. We coupled our phase noise model to a linear filtering cavity model and studied the cavity's effectiveness in reducing the pulse-to-pulse phase jitter to an acceptable level for XUV spectroscopy. We proposed and numerically validated a non-destructive quantum logic-type readout scheme for trapped ions outside the Lamb-Dicke regime and presented a technique to circumvent both first-order Doppler broadening and the photon recoil shift for trapped ion spectroscopy carried out with single-sided excitation. Our numerical study demonstrates the feasibility of trapped ion XUV spectroscopy, including weak transitions. 

We considered the special case of XUV Ramsey-comb spectroscopy (RCS) on He$^{+}$ co-trapped and read out with a Be$^{+}$ logic ion. We developed a model for the motional excitation dynamics of RCS of single trapped ions outside the Lamb-Dicke regime which could be coupled to the phase noise and filtering cavity models. This allowed us to realistically simulate extreme-ultraviolet RCS of the $1S-2S$ transition in He$^{+}$ using both our proposed logic-type readout scheme and a destructive state-selective ionization readout scheme. We find that a measurement with $\sim$ 10 kHz precision is feasible with both techniques within reasonable data acquisition times when combined with the use of a filtering cavity with finesse around 4000 for noise suppression. The quantum logic readout technique, however, allows for a dramatic decrease in total data acquisition time with increasing excitation probability, ultimately allowing for a higher precision measurement. A precision measurement of the $1S-2S$ transition in He$^{+}$, enabled by the techniques discussed in this article, would allow for new tests of quantum electrodynamics or an improved determination of fundamental constants such as the Rydberg constant, the alpha particle and helion charge radii~\cite{krauth2019paving,Grundeman}. 

\begin{acknowledgments}
\textit{Acknowledgements.} K.E. acknowledges support from the Dutch Research Council (NWO) through grant 16MYSTP and OCENW.XL21.XL21.122. \\

\textit{Data availability.} Data are available from the authors upon reasonable request.
\end{acknowledgments}

\section{Appendix}

\subsection{Experimental setup}
The RCS laser and vacuum system are schematically shown in Fig. \ref{fig:experimental_setup} below. 

\begin{figure*}
\includegraphics[width=\textwidth]{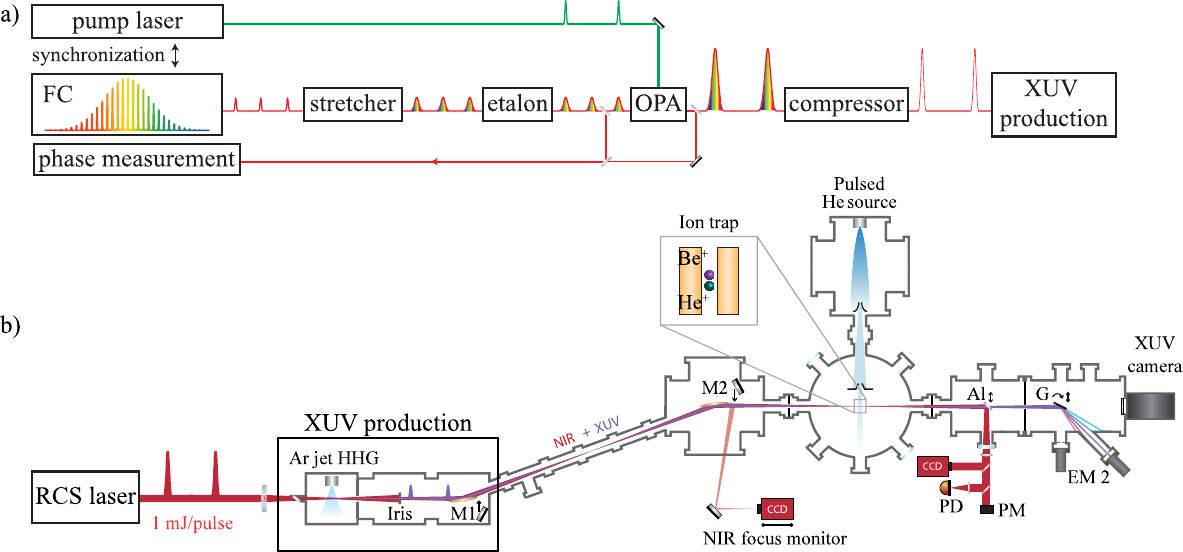}
\caption{\label{fig:experimental_setup} a) RCS laser system. A double-pulse pump laser at 532 nm is repetition-rate locked to an ultrastable frequency comb laser (FC) that itself is locked to an atomic clock. The FC pulses are stretched and spectrally clipped in a 4-f grating stretcher after which they are sent through a filtering cavity (etalon) to reduce its optical phase noise. The filtered pulse train is then amplified in three stages in an LBO-based optical parametric amplifier (OPA). After amplification, the pulses are recompressed in a grating compressor to $\sim$ 150 fs with energies of a few mJ per pulse that can be focused to peak intensities on the order of $10^{14}$ W/cm$^{2}$. After the compressor, the pulses are sent to the XUV production chamber shown in b). b) The He$^{+}$ spectroscopy vacuum setup. Two selectively-amplified pulses are focused into a beam of argon atoms where XUV radiation is produced by HHG. The NIR+XUV pulse is focused onto the He$^{+}$ ion in the ion trap using a one-to-one telescope comprised of a pair of grazing incidence, gold-coated toroidal mirrors (M1 and M2). An aluminum filter (Al) after the spectroscopy chamber reflects wavelengths above 60 nm and is used to couple out the NIR beam. An aluminum grating (G) can be used to disperse the harmonics onto an electron multiplier (EM). An Andor Newton charge-coupled device (CCD) camera (XUV camera) is used to monitor the XUV beam.}
\end{figure*}

\subsection{TDSE simulations of RCS on He$^{+}$}\label{TDSE_simulations_appendix}

In a previous publication \cite{Grundeman} on the excitation of the $1S-2S$ transition in He$^{+}$ with a single pulse, we used Time-dependent Schr\"odinger Equation (TDSE) simulations to gain insight in the excitation process. Here we do the same for two pulses to simulate the RCS excitation process. We model a single He$^+$ ion initially in its ground state and expose it to an electric field consisting of a near-infrared (NIR) $\lambda=790$ nm laser pulse and its odd harmonics. To simulate a Ramsey-fringe, the time delay between two pulses is linearly increased in order to sample the excited state population at eight interpulse delays $\Delta t$ per Ramsey fringe. Although the minimum delay between two RSC pulses corresponds to the 8 ns RCS laser repetition period, we separate our pulses by an initial macrodelay of $600$ fs (see eq. (\ref{delays})) to make the simulation computationally tractable, while the separation is large enough so that no direct interference between the two light pulses occurs. Each of the two time-separated NIR pulses has a Gaussian time envelope with a full-width at half-maximum of $150$ fs. The polarization of the NIR and XUV fields is assumed to be linear and parallel to the quantization axis along $z$. 

We solve the time-dependent Schr\"odinger equation for a ground-state He$^+$ ion given by
\begin{equation}\label{electronic excitation hamiltonian}
i\hbar\frac{\partial \Psi ({\bf r},t)}{\partial t} = 
\left[H_{at} + H_{int}(t)
\right]\Psi ({\bf r},t),
\end{equation}
where $\hbar$ is the reduced Planck constant and where $H_{at}$ is the 3D atomic Hamiltonian
\begin{equation}\label{H atomic}
H_{at}=\frac{p^2}{2m}-\frac{2e^2}{4\pi\epsilon_{0}\vec{r}} 
\end{equation}
with $p$ the momentum operator, $m$ the electron mass, $e$ the electron charge, $\epsilon_{0}$ the permittivity of free space, and $\vec{r}$ the electron position vector with respect to the nucleus. 

The atom-field interaction 
Hamiltonian $H_{int}$ in the velocity gauge is given by
\begin{equation}\label{H interaction}
H_{int}(t) =  -e p_{z} A(t)
\end{equation}
where $A(t)$ is the vector potential derived from the electric field $E(t)$ as 
\begin{equation}
E(t)=-\frac{\partial A(t)}{\partial t}
\end{equation}
and where $p_{z}$ the linear momentum along the $z$ direction. Time propagation of the TDSE is carried out by a 
standard Peaceman-Rachford scheme coupled to an inverse iteration procedure,
very similar to the one previously derived by Kulander and Schafer, and Ta\"ieb \textit{et al.} (see \cite{Schafer_Kulander, Taieb_96}). 

\begin{figure*}
\includegraphics[width=0.65\textwidth]{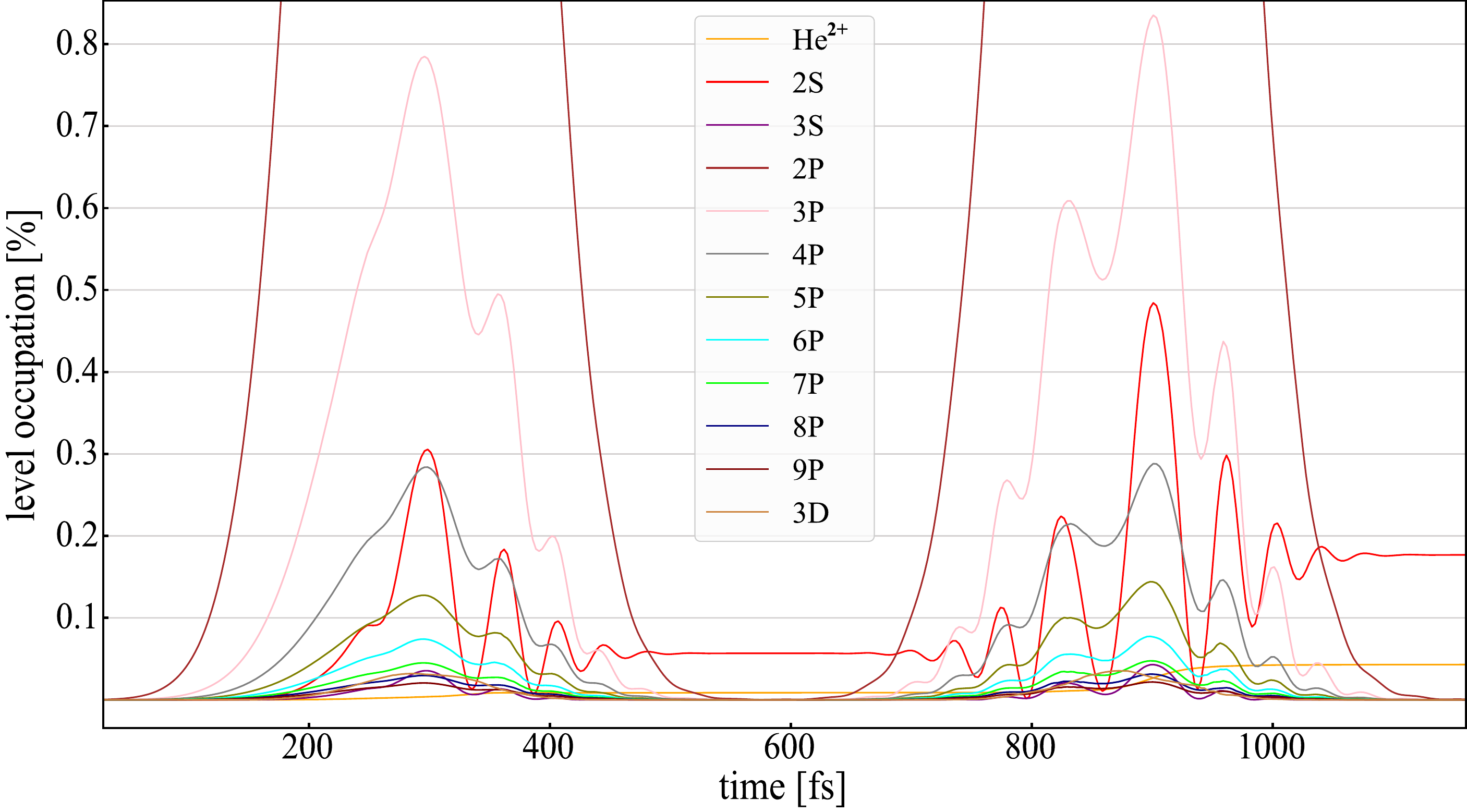}
\caption{\label{fig:level_over_time_fig} Projection of the electron wavefunction onto the unperturbed eigenstates of $H_{at}$ throughout the RCS excitation. Although the wavefunction acquires a very strong 2P character throughout the excitation, only population in the 2S and in the doubly-ionized state remain after the RCS excitation. The level occupations feature strong oscillations in all but the 2P state. The 2S and nP states for $n \neq 2$ oscillate in phase with one another, and are anti-correlated ($\pi$ out of phase) with the 1S level, indicating that all higher-lying states are reached via the 2S state. When we varied the intensity of the excitation fields, the oscillation periods seen in the signal remained largely the same, indicating that they do not originate from a simple ac-Stark effect.}
\end{figure*}

A measure for the excitation of the different atomic states during and after the RCS pulses is obtained by projecting the wavefunction $\Psi ({\bf r},t)$ onto the eigenstates of the unperturbed He$^+$ Hamiltonian $H_{at}$. An interesting aspect of these simulations is that we can get insights into the excitation dynamics throughout the RCS excitation in a way that would be impossible experimentally. Fig. \ref{fig:level_over_time_fig} shows projections of the He$^{+}$ electronic wavefunction onto the unperturbed eigenstates of the atomic Hamiltonian. These curves should not be interpreted as real populations; Only the final result after excitation gives the true populations. Rather, we take them to be indicative of the character of the wavefunction throughout the RCS excitation. The 2P character is significant while the laser field is present but quickly decays in the absence of the field. The 2S character shows coherent build-up over the Ramsey sequence and displays an oscillation during the presence of the laser field which couples to all other excited states, suggesting that the 1S state only couples to these higher states via the 2S state. We investigate a possible intensity-dependence to the oscillation frequency exhibited by the 2S and higher lying states by determining the peak-to-peak oscillation frequency of the populations for various intensities. We find no straightforward dependence on the intensity that could point to, e.g. a linear ac-Stark shift being responsible for these oscillations. 

Fig. \ref{fig:level_over_time_fig} shows the evolution and final state of the He$^{+}$ atom for one particular delay between the two excitation pulses. We then simulated the 2S excitation (or He$^{2+}$ ion generation) for multiple delays, and constructed the simulated Ramsey fringe as shown in Fig. \ref{fig:fringes_fig}. The very small excitation to the 2P is neglected because the minimum pulse delay of about 200 ns in real RCS on He$^{+}$ is much longer than the decay time from the first few lower P states.

\subsection{Ion wave function after RCS sequence}\label{appendix_ion_derivation}
In this section, we give more details on the derivation of the RCS signal in the case of trapped ions outlined in section \ref{RCS on trapped ions}. Excitation of He$^{+}$ in a trap by the 32 nm and 790 nm (both coming from the same side) leads to a large momentum kick to the ion. This shows itself as a distribution of trap states, which can form a so-called coherent state. We model the effect of the first RCS pulse on the ion wavefunction using the RCS operator $\hat{\mathcal O }$ defined in eq. (\ref{RCS_operator}) as $\hat{\mathcal O } = \lambda_{2}\hat{D}(\alpha)\otimes \hat{T}+\lambda_{1}\mathds{1}$, where $\lambda_{2}$ is the 2S excitation amplitude, taken as the square root of the $10^{-3}$  $1S-2S$ electronic excitation probability. $\lambda_{1}$ is the square root of the probability to remain in the electronic ground state, which we approximate as $P_{g} = \lambda_{1} =1$ in the simulations. This means that the second RCS pulse interacts with the full (amplitude 1) ground state wavepacket, which amounts to a negligible overestimation of 0.05\% on its amplitude. $\hat{D}(\alpha)$ is the displacement operator for a displacement of amplitude $\alpha =\eta e^{i\xi}$, with $\eta$ the Lamb-Dicke parameter and $\xi$ the angle of displacement in complex phase space (in our case $\xi=\pi/2$, representing pure momentum displacement).

Acting with this operator on the ground state wave function at time $t_{0}$ gives: 
\begin{equation}\label{after_P1}
    \Tilde{\psi}(t_{0}) = \hat{\mathcal O }\psi(t_{0})
     = \lambda_{2}\sum_{n,l}^{\infty}\beta_{n}C_{n,l}\ket{l}\otimes \ket{\uparrow} \\
    +\sum_{n}\beta_{n}\ket{n}\otimes \ket{\downarrow}
\end{equation}
with $\ket{\downarrow}$ the electronic ground state (1S) and $\ket{\uparrow}$ the electronic excited state(2S), and the motional state as given in eq. (\ref{thermal dist}), making use of the identity $\sum_{n=0}^{\infty}\ket{l}\bra{l} = \mathds{1}$, and where the displacement operator matrix element $C_{n,l}$ is given by \cite{Wunsche}:
\begin{multline}\label{3.11}
C_{n,l} = \bra{l}D(\alpha)\ket{n} = e^{-\frac{1}{2}|\alpha|^{2}} \times \\
\begin{cases}
     \left(\frac{l!}{n!}\right)^{\frac{1}{2}}(-\alpha^{*})^{n-l}\mathscr{L}^{n-l}_{l}(|\alpha|^2) &
\textrm{for } n\geq l  \\
 \left(\frac{n!}{l!}\right)^{\frac{1}{2}}\alpha^{l-n}\mathscr{L}^{l-n}_{n}(|\alpha|^2) & \textrm{for } l> n
    \end{cases}
\end{multline}

\noindent  where $\mathscr{L}^{j}_{k}(x)$ are the generalized Laguerre polynomials. $\Tilde{\psi}(t_{0})$ is then propagated to time $t^{\prime}$ corresponding to the RCS interpulse delay, using the time evolution operator $\hat{U}(t^{\prime})$ (see eq.(\ref{time_evolution_op})). We can ignore the phase factor of the motional ground state energy $\hbar \omega_{\textrm{sec}}/2$, with $\omega_{\textrm{sec}}$ the secular frequency because it is common to all terms. At the time of the second pulse, $t'$, we obtain for the wavefunction (due to the first excitation pulse) the following result:
\begin{multline}\label{ion_wave_function_evolution}
    \Tilde{\psi}(t^{\prime}) = \lambda_{2}\sum_{n,l}^{\infty}\beta_{n}C_{n,l}e^{-i(\omega_{sec,l}+\omega_{tr})t^{\prime}}\ket{l}\otimes \ket{\uparrow} \\
    +\sum_{n}\beta_{n}e^{-i\omega_{sec,n}t^{\prime}}\ket{n}\otimes \ket{\downarrow}
\end{multline}
\noindent where $\omega_{sec,l} = \omega_{\textrm{sec}} \times l$, $\omega_{sec,n} = \omega_{\textrm{sec}} \times n$. The RCS operator $\hat{\mathcal O }$ is then applied on $\Tilde{\psi}(t^{\prime})$, yielding the ion wavefunction after the two pulse sequence as
\begin{multline}\label{RCS_ion}
\psi(t^{\prime})_{RCS} =
\lambda_{2}\sum_{n}\beta_{n}\sum_{l}C_{l,n}e^{-i(\omega_{sec,l}+\omega_{tr})t^{\prime})}\ket{l}\otimes\ket{\uparrow} \\
+ \lambda_{2}\sum_{n}\beta_{n} e^{-i(\omega_{sec,n}t^{\prime} +\Delta(t^{\prime}))}\sum_{k}C_{k,n}\ket{k}\otimes\ket{\uparrow} \\
\end{multline}
where a relative phase term $\Delta\phi(t^{\prime}) = \phi_{CEO}+\delta\xi(t^{\prime})$ has been included to account for the carrier-envelope phase slip between RCS pulse 1 and 2, and a relative phase noise term $\delta\xi(t^{\prime})$. In the simulation, the amplitude of both terms in equation \ref{RCS_ion} is reduced by the probability amplitude of ionization of He$^{+}$(2S). The first and second terms of eq. (\ref{RCS_ion}) represent the contribution to the excited state created by the first and second RCS pulses, respectively. Projecting eq. (\ref{RCS_ion}) onto the electronic excited state (summed over all motional states $\ket{\psi_{2S}}=\ket{m}\otimes\ket{\uparrow}$) gives the probability amplitude for finding the ion in the electronic excited state as
\begin{multline}
 \bra{\psi_{2S}}\ket{\psi(t^{\prime})_{RCS}} = 
 \lambda_{2}\sum_{n,m}\beta_{n}C_{m,n} \\ \times
 \left (e^{-i(\omega_{sec,n}t^{\prime}+\Delta\phi(t^{\prime}))} + e^{-i(\omega_{sec,m}+\omega_{tr})t^{\prime}}\right)
\end{multline}
making use of the orthogonality of motional and electronic states $\bra{m}\ket{n}=\delta_{m,n}$. The RCS signal is then given by 
\begin{equation}
RCS_{\textrm{signal}} = |\bra{\psi_{2S}}\ket{\psi(t^{\prime})_{RCS}}|^{2}
\end{equation}
which features cross terms from both the exponentials in the parenthesis and from products of the form $\beta_{n}\beta_{m}^{*}$ in equation (\ref{RCS_prob_amp}) (thereby making it very cumbersome to write out the full expression in the present article).

\subsection{Monte Carlo simulation details}
Based on the formalism presented in \ref{RCS on trapped ions}, we perform Monte Carlo simulations of the motional excitation of a trapped He$^{+}$ ion interrogated with the RCS technique (see \ref{RCS_technique}). Each iteration of the simulation initializes an ion with a motional state distribution belonging to a temperature $T$ described by equation (\ref{thermal dist}) and randomly selects a value from the phase noise set $\delta \xi(t)$ generated as described in subsection \ref{laser phase noise} for NIR pulses at 790 nm.

Using this simulated noise data, we obtain the probability of finding a He$^{+}$ ion in the 2S state. After each of the two RCS pulses, the amplitude of the contribution to the 2S state is reduced by the probability of ionization from He$^{+}(2S)$ to He$^{2+}$.  We take quantum projection noise into account by randomly sampling a flat distribution with values between zero and one at the end of every transition amplitude calculation. If the value produced by this random sampling is below the calculated electronic transition probability, we consider it a successful count and add it to the accumulated signal, mimicking the random sampling of the quantum mechanical probability distribution.

\newpage

\nocite{*}

\clearpage
\bibliography{bibliography}

\end{document}